\documentclass{ieeeaccess}

\usepackage{cite}
\usepackage{amsmath,amssymb,amsfonts}
\usepackage{algorithmic}
\usepackage{graphicx}
\usepackage{textcomp}

\usepackage{multirow} 
\usepackage[table]{xcolor} 
\usepackage{booktabs}

\def\BibTeX{{\rm B\kern-.05em{\sc i\kern-.025em b}\kern-.08em
    T\kern-.1667em\lower.7ex\hbox{E}\kern-.125emX}}
\begin{document}
\history{}
\doi{}

\title{Reconfigurable Computing Applied to Latency Reduction for the Tactile Internet}
\author{\uppercase{Jos\'{e} C. V. S. Junior}\authorrefmark{1},
\uppercase{Matheus F. Torquato\authorrefmark{2}, Toktam Mahmoodi\authorrefmark{3},}
\uppercase{Mischa Dohler\authorrefmark{3}, \IEEEmembership{Fellow, IEEE} and Marcelo A. C. Fernandes\authorrefmark{1,4,5}} }

\address[1]{Laboratory of Machine Learning and Intelligent Instrumentation (LMLII), nPITI-IMD, Federal University of Rio Grande do Norte, 59078-970, Natal, Brazil.}

\address[2]{College of Engineering, Swansea University, Swansea, Wales SA2 8PP, UK.}

\address[3]{Centre for Telecommunications Research, Department of Engineering, King’s College London, London WC2R 2LS, UK.}

\address[4]{Department of Computer and Automation Engineering, Federal University of Rio Grande do Norte, \mbox{Natal, 59078-970, Brazil.}}

\address[5]{(Current address) John A. Paulson School of Engineering and Applied Sciences, Harvard University, 02138, Cambridge, MA, USA.}


\tfootnote{This study was partially financed by the Coordena\c{c}\~ao de Aperfei\c{c}oamento de Pessoal de N\'ivel Superior (CAPES)---Finance Code 001.}

\markboth
{}
{}

\corresp{Corresponding author: Marcelo A. C. Fernandes (mfernandes@dca.ufrn.br or macfernandes@seas.harvard.edu ).}

\begin{abstract}
Tactile internet applications allow robotic devices to be remotely controlled over a communication medium with an unnoticeable time delay. In a bilateral communication, the acceptable round trip latency is usually in the order of $1 \, \text{ms}$ up to $10 \, \text{ms}$ depending on the application requirements. It is estimated that $70\%$ of the total latency is generated by the communication network, and the remaining $30\%$ is produced by master and slave devices. Thus, this paper aims to propose a strategy to reduce the $30\%$ of total latency that is produced by such devices. The strategy is to apply reconfigurable computation using FPGAs to minimize the execution time of device-associated algorithms. With this in mind, this work presents a hardware reference model for modules that implement nonlinear positioning and force calculations as well as a tactile system formed by two robotic manipulators. In addition to presenting the implementation details, simulations and experimental tests are performed in order to validate the proposed model. Results associated with the FPGA sampling rate, throughput, latency and post-synthesis occupancy area are analyzed.
\end{abstract}


\titlepgskip=-15pt

\maketitle

\section{Introduction}
\label{sec:introduction}

\PARstart{T}{he} Tactile internet is conceptually defined as the new generation of internet connectivity which will combine very low latency with extremely high availability, reliability and security \cite{tactile_basic2015_Mischa}. Another feature pointed out is that this new generation will be centered around applications that use human-machine communications (H2M) alongside devices that are compatible with tactile sensations \cite{tactile_network2015_RealizingTactile, 2017_challengesHapticTactile}.

A tactile internet environment is basically composed of a local device (known as a master) and a remote device (known as a slave), where the master device is responsible for controlling the slave device over the internet through a two-way data communication network \cite{tactile_basic2016_Martin} \cite{tactile_basic2016_The5G}. Bidirectional communication is needed to simulate the physical laws of action and reaction, where action can be represented as sending operational commands and reaction can be represented as the forces resulting from that action. In tactile internet applications, the desired time delay for device communication is characterized by an ultra-low latency.  In bilateral communication, the required round trip latency ranges from $1 \, \text{ms}$ up to $10 \, \text{ms}$ depending on the application requirements \cite{REF_TIME_MS_01, 8399482, REF_TIME_MS_02, tactile_network2016_5GEnable}.

According to \cite{tactile_network2015_TowardsLatency}, it can be noticed that in a tactile internet application, $30\%$ of the total system latency is generated by the master and slave devices. These devices demand high processing speeds as repeated execution of a variety of computationally expensive algorithms and techniques are required. These algorithms involve the use of arithmetic operations and calculations of linear and nonlinear equations that need to be computed at high sampling rates in order to maintain application fidelity. The remaining $70\%$ of the latency is caused by the communication network, which makes them unsuitable for such latency constraints \cite{internetSkills_mischa2017}. To minimize this problem, some research groups have been studying prediction techniques, where many algorithms have been studied and proposals using artificial intelligence (AI) have proved to be effective \cite{predicaoIAFPGA2015}. On the other hand, the implementation of complex AI-based prediction methods can further increase the latency of the computer systems present in master and slave devices.

Alternatively, new approaches such as reconfigurable computing can improve the performance of master and slave devices in a tactile system environment. Reconfigurable computing with field-programmable gate arrays (FPGAs) enables the creation of customizable hardware which allow algorithms to be parallelized and optimized at the logical gate level to speed up their operations. Literature results show that computationally expensive algorithms can achieve speedups of up to $1000\times$ over software implementations when custom-implemented in FPGAs \cite{marceloAlisson2014, 8626462, 8678408, Torquato2019, 8574886, electronics8060631, NORONHA2019138}.

In this context, this paper proposes an implementation to target reducing the $30\%$ of the total latency related to tactile devices. The project uses reconfigurable computation in FPGA to minimize the execution time of algorithms associated with master and slave devices. The use of reconfigurable computing allows the parallelization of algorithms and latency reduction compared to software systems embedded in traditional architectures with general purpose processors and microcontrollers. In an effort to validate the  proposed strategy, this paper presents a discrete reference model that can be adjusted for different types of master and slave devices in a tactile internet system. Validation results, throughput, and post-synthesis figures obtained for the proposed hardware implementation using FPGA reconfigurable computing are presented. Comparisons with other works in the literature show that the use of reconfiguration computing can significantly accelerate the processing speed in tactile devices.

\section{Related Work}

The authors of \cite{2018_tactilePhysicalSystemDesign} presented a tactile internet environment that used a glove type device in conjunction with a robotic manipulator. The environment was developed using a general purpose processor, which made the execution of the algorithms sequential. In order to send the data, the tactile glove produced a latency of approximately $4.82 \, \text{ms}$, and the hardware responsible for performing the inverse kinematics calculations took an interval of $0.95 \, \text{ms}$. The latency values obtained in this application could be improved by hardware structures that allow algorithms parallelization. 

Studies in the literature demonstrate the benefit of using FPGA to accelerate the sample rate for data acquisition from devices associated with haptic systems. The authors of \cite{reduction_fpga2009_ImprovedHaptic} presented an implementation for controlling a 3-DoF (Degree of Freedom) device. The presented technique proposed to increase the device sampling rate using FPGA hardware together with a real-time operating system (RTOS) in order to increase the resolution acquisition of the stiffness sensor. The control technique presented was developed in $32$-bit fixed point, and trigonometric functions were implemented using lookup tables.

The work described in \cite{fpga_haptic2009_HapticComunnication} presented a control system for one-dimensional haptic devices (1-DoF). The FPGA control implementation used single-precision floating point representation (IEEE std 754) and the algorithms performed all calculations in $50 \mu \text{s}$. The processing time was satisfactory; however, the data frame size to be sent over the network increased with the size of the DoF. This peculiarity can increase latency for more complex haptics systems with many DoFs. In the same topic of previous works, an implementation for bilateral control of single-dimensional haptic devices (1-DoF) was presented in \cite{reduction_fpga2013_AStudyFpga}. A more accurate control techniques based on the sliding mode control (SMC) was implemented in FPGA, and to assist in performing the complex calculations, the CORDIC (COordinate Rotation DIgital Computer) was used. The hardware was designed to locally control two devices, one master and one slave. In the implementation, a $24$-bit fixed point was used, of which $9$ bits in the integer part and $14$ bits for the fractional, and the total execution time of the controllers was of $7.2375 \mu \text{s}$.

The works \cite{reduction_fpga2009_ImprovedHaptic}, \cite{fpga_haptic2009_HapticComunnication} and \cite{reduction_fpga2013_AStudyFpga} presented a control that depends directly on the encoder reading of the device motors. Usually in commercial models, accessing the device electronics can be tricky requiring some reverse engineering and specific knowledge to make the appropriate encoder connections. On the other hand, some works abstract the data acquisition and work directly with robotics algorithms. These algorithms may require high computational power that can surpass the capabilities of many general-purpose processors (GPPs) that perform the operations sequentially.

Some studies demonstrate the benefit of using FPGA to accelerate robotic manipulation algorithms related to haptic systems. A hardware architecture implemented in FPGA for performing the forward kinematics of 5-DoF robots using floating point arithmetic was described in \cite{2010_fpgaKinematicRoboticManupulator}. In this hardware implementation all the forward kinematics calculations were performed within $1.24 \mu s$ which represents $67$ clock cycles in a frequency of $54 \, \text{MHz}$. The equivalent software implementation has a total processing time of $1.61036 \, \text{ms}$. Overall, the hardware implementation is $1298\times$ faster than the software implementation, which means a considerable acceleration in the forward kinematics processing time.

The authors of the paper \cite{2013_fpgaParallelRobotKinematic} presented an FPGA implementation of inverse kinematics, velocity calculation and acceleration of a $3$-DoF robot. Three systems were created: the first one did not use any arithmetic co-processor and floating point operations were performed in software; in the second system a floating point co-processor was used which allowed the execution of the four basic mathematical operations in hardware; lastly, the third system also had a custom arithmetic co-processor but in this case it allowed hardware computation of square root. The overall times to perform the calculations were $ 2324 \mu \text{s}$, $560 \mu \text{s}$ and $143 \mu \text{s}$ and the total logic elements used from the entire device were $4501$ ($4\%$), $5840$ ($5\%$) and $7219$ ($6\%$), respectively. The work uses hardware-software to implement inverse kinematics, in which critical parts were implemented in FPGAs to accelerate the whole process.

In \cite{fpga_2014_fixedPointKinematics} is presented a hardware to control a $6$-DoF device using $32$-bit fixed point representation, where $21$ bits were used for the fractional part and $11$ bits for the integer part. In that work, a CORDIC implementation was used to assist in performing the trigonometric calculations. The total time spent to compute the forward kinematics was $3 \mu \text{s}$ and for the inverse kinematics the time was $4.5 \mu \text{s}$ for a clock of $50 \, \text{MHz}$. However, in the presented proposal, some calculations were performed sequentially, that is, for the execution of the forward kinematics it was necessary $150$ clock cycles and for the inverse, $225$ cycles. The use of partial parallelization in the execution of robotic manipulation algorithms provided a significant increase in system throughput. Nevertheless, it is important to note that there is still room for improvement since all calculations can be computed in parallel.

Another hardware implementation of inverse kinematics was presented in \cite{2013_fpgaKinematicsCordic}. The device used was a 10-DoF biped robot. A CORDIC implementation was used to perform the trigonometric calculations. The execution time needed to compute the kinematics of the $10$ joints in FPGA was of $0.44 \mu \text{s}$. In this paper, a comparison with a software implementation was also performed, and the time taken to perform the same calculations was $3342 \mu s$, i.e. the gain on execution, or speedup, on custom FPGA hardware was $7595 \times$. The resulting error between both implementations was acceptable for this specific control.

In \cite{2015_fpgaKinematicArticuledRobot} it was presented an FPGA implementation of the forward and inverse kinematics of a $5$-DoF device. The hardware was developed using a fixed point representation where $32$ bits were used for the angles representation and $15$ bits for the fractional part. For the device spatial positioning, $16$ bits were used of which $7$ bits for the fractional part. In the implementation of trigonometric functions, a combination of techniques using lookup tables (LUTs) and Taylor series was used. To perform the necessary calculations, a finite-state machine model (FSM) was used to reduce the use of hardware resources, however, the use of such FSM generated a sequential computation of the robotic manipulation algorithms. In this model, the forward kinematics implementation achieved a runtime of $680 \, \text{ns}$ and the inverse $940 \, \text{ns}$, that is, for the $50 \, \text{MHz}$ clock, the forward kinematics took $34$ clock cycles and the inverse kinematics took $47$ cycles. Using such approaches to reduce the use of hardware resources increases computation runtime. For tactile device applications, it is important to optimize the runtime rather than the use of hardware resources.

Similarly, an FPGA implementation of forward and inverse kinematics for a $7$-DoF device was presented in \cite{2017_fpga_SurgicalRobotKinematic}, however, only $3$-DoF required to control the device movement were implemented in hardware. The proposal used a $32$-bit fixed point representation and a CORDIC was used to execute the trigonometric functions. To validate the proposal, the FPGA was set to receive the three reference angles, perform the forward kinematics and then the inverse. The model was developed based on pipeline and the operating frequency used was of $100 \, \text{MHz}$. As a result, the model calculation took $2 \mu \text{s}$ to perform the entire kinematics algorithm, which represented $200$ clock cycles.

In this context, it is possible to realize that the use of reconfigurable FPGA-based computing can accelerate haptic device control algorithms. Unlike traditional hardware that processes information sequentially, FPGA enables parallel information processing. However, most studies from the literature have developed partially parallel implementations, that is, implementations in which parts of the used algorithms are executed sequentially. Unlike the researches previously mentioned, this study presents a new approach in which the execution of the robotic manipulation algorithms are performed in a full-parallel hardware implementation. This proposed implementation provides a latency reduction for the tactile devices and enables tactile internet applications.

\section{Discrete Model of the Tactile Internet} \label{sec:GenericTactileModel}

A discrete model of the tactile internet system is proposed and presented in Figure \ref{fig:Fig_TactileModel_Generic}. This model consists of seven subsystems: the Operator (OP), Master Device (MD), Hardware of the MD (HMD), Network (NW), Hardware of the SD (HSD), Slave Device (SD) and the Environment (ENV). It is assumed that the signals are sampled at time $t_s$.

\Figure[ht]()[width=\textwidth]{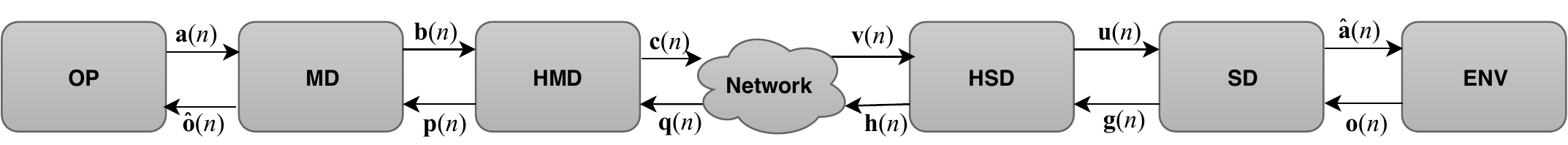}
{Proposed discrete model of the tactile internet system.\label{fig:Fig_TactileModel_Generic}}

The OP is an entity responsible for generating stimuli that can be in the form of position signals, speed, force, image, sound or any other. These stimuli are sent to the devices involved so that some kind of task can be performed in some kind of environment. The environment, the ENV subsystem, receives the stimuli from the OP and generates feedback signals associated with sensations such as reactive force information and tactile information that are sent back to the OP. The interaction between the OP and the ENV is performed through the master and slave devices, MD and SD, respectively.

Specifically in this work, MD is characterized as a local device, SD as remote one and both of them are responsible for transforming the stimuli and sensations associated with OP and ENV into signals to be processed. Tactile devices (MD and SD) can take the form of robotic manipulators, haptic devices, tactile gloves and others that may be developed in the future. In the coming years, the introduction of new types of sensors and actuators is expected that will form the basis for the development of new tactile devices.

Although there are no tactile internet standards nor products yet, it can be affirmed that future tactile devices will be integrated with a hardware responsible for all operational metrics and calculations. Within this conjecture, this work adds a couple of modules to the discrete model (as per Figure \ref{fig:Fig_TactileModel_Generic}), called HMD and HSD. HMD is responsible for performing all transformations and calculations associated with MD, and HSD performs the equivalent operations for the SD. Several algorithms associated with transformation, compression, control, prediction will be under the responsibility of these two modules.

Based on the model presented in Figure \ref{fig:Fig_TactileModel_Generic}, the signals generated by the OP can be characterized by the array $\mathbf{a}(n)$ expressed as
\begin{equation}
\mathbf{a}(n) = \left[a_1(n),\dots, a_i(n), \dots, a_{N_{OP}}(n)\right],
\end{equation}
where $a_i(n)$ is the $i$-th stimulus at the $n$-th instant and $N_{OP}$ is the total number of stimuli signals generated by the OP. At every $n$-th moment the stimulus array, $\mathbf{a}(n)$, is received by the MD which transforms the stimuli into a set of $N_{MD}$ signals expressed as
\begin{equation}
\mathbf{b}(n) = \left[b_1(n), \dots, b_i(n), \dots, b_{N_{MD}}(n)\right],
\end{equation}
where $b_i(n)$ is the $i$-th signal generated by MD at the $n$-th instant. It can be stated that at each $n$-th moment a set of stimuli $\mathbf{a}(n)$ generates a set of signals $\mathbf{b}(n)$ that depend on the type of MD and the sensor set associated with the device. Especially important is the fact that the signals generated by MD, $\mathbf{b}(n)$, have heterogeneous characteristics in which each $i$-th signal $b_i(n)$ can represent an angle, spatial coordinate, pixel of an image, audio sample or any other information associated with a stimulus generated by OP. In practice, the signals grouped by the $\mathbf{b}(n)$ array originate from sensors coupled to the MD and the amount of data may vary according to the amount of information to be sent, $N_{MD}$.

The set of signals, expressed by $\mathbf{b}(n)$ are sent to the HMD (Figure \ref{fig:Fig_TactileModel_Generic}) which has the function of processing this information before sending it to the NW subsystem. Calculations associated with calibration, linear and nonlinear transformations and signal compression are performed by the HMD. Essentially the majority of the computational effort of MD is in this subsystem. At each $n$-th instant $t_s$ the HMD processes the array $\mathbf{b}(n)$ generating an information array $\mathbf{c}(n)$ expressed by
\begin{equation}
\mathbf{c}(n) = \left[c_1(n), \dots, c_i(n), \dots, c_{N^f_{HMD}}(n)\right],
\end{equation}
where $c_i(n)$ is the $i$-th signal generated by HMD towards the subsystem NW at the $n$-th instant $t_s$ and $N^f_{HMD}$ is the numbers of signals.
$N^f_{HMD}<N_{MD}$ is expected to minimize latency during the transmission in the NW subsystem.

The NW subsystem, as shown in Figure \ref{fig:Fig_TactileModel_Generic}, characterizes the communication medium that links OP to ENV. In this model, the data propagates through two different channels called the forward channel, that transmits the OP data towards the ENV, and the backwards channel, that transmits the ENV signals towards the OP. The signal transmitted by the forward and backwards channels may be disturbed and delayed. In the case of the forward channel, the received signal, $\mathbf{v}(n)$, may be expressed as
\begin{equation}
\mathbf{v}(n) =\left [v_1(n), \dots, v_i(n), \dots, v_{N^f_{HMD}}(n)\right],
\end{equation}
where
\begin{equation}\label{EQChannel1}
v_i(n) = c_i \left(n-d_i^f(n) \right) + r^f_i(n)
\end{equation}
in which, $r^f_i(n)$ represents the added noise and $d_i^f(n)$ represents a delays associated with the $i$-th information sent in $\mathbf{c}(n)$.
In this model, the noise can be characterized as a random Gaussian variable of zero mean and $\sigma^2_{rf}$ variance, and the delays are characterized as integers, that is, they occur at a granularity of $t_s$. It is important to note that the NW subsystem can take the shape of the Internet, a metropolitan network (MAN), a local area network (LAN), or even a direct connection between an MD and a workstation or computer.

As shown in Figure \ref{fig:Fig_TactileModel_Generic}, the HSD receives the $\mathbf{v}(n)$ signal through the forward channel and has the role of generating control signals to the SD through the signal
\begin{equation}
\mathbf{u}(n) =\left[u_1(n), \dots, u_i(n), \dots, u_{N^f_{HSD}}(n)\right],
\end{equation}
 where $N^f_{HSD}$ is the number of control signals and $u_i(n)$ is $i$-th control signal at the $n$-th instant $t_s$ associated with the array $\mathbf{u}(n)$. It is important to note that there may be various types of SD: from real robotic handlers to virtual tools in computational environments. Thus, it can be stated without loss of generality that HSD can perform an inverse processing to HMD in addition to specific algorithms associated with the type of SD. For example, if the SD is a robotic handler, HSD must additionally implement closed loop control algorithms, whereas if SD is a virtual arm HSD must implement positioning algorithms for a given virtual reality platform. SD does not have to correspond directly with MD, e.g. MD can be a glove while SD is a drone. However, it is desirable that the stimulus generated by the SD is a copy of the stimulus generated by the OP, that is, within the model presented in Figure \ref{fig:Fig_TactileModel_Generic}, it can be understood that SD generate a signal expressed as
\begin{equation}
\mathbf{\hat{a}}(n) = \left[\hat{a}_1(n),\dots, \hat{a}_i(n), \dots, \hat{a}_{N_{OP}}(n)\right],
\end{equation}
where $\hat{a}_i(n)$ is an estimate of the $i$-th stimulus $a_i(n)$ generated by the OP. Thus, the estimate of the stimulus generated by OP, $\hat{a}_i(n)$, is applied to the ENV subsystem representing a given real or virtual environment in which OP is interacting.

In the backwards direction, the stimulus actions generated by OP, $\mathbf{a}(n)$, and represented by $\mathbf{\hat{a}}(n)$, receives a group of reactions from the ENV subsystem that can be characterized in the model by the set of signals expressed by
\begin{equation}
\mathbf{o}(n) = \left[o_1(n), \dots, o_i(n), \dots, o_{N_{ENV}}(n)\right],
\end{equation}
where $N_{ENV}$ is the number of stimulus signals and $o_i(n)$ is $i$-th stimulus signal at the $n$-th instant $t_s$. Reaction signals grouped into $\mathbf{o}(n)$ can be in the form of strength, touch, temperature, etc.

Reaction signals are captured by the SD that turns this information into electrical signals from real or virtual sensors, if the SD is in a virtual reality environment. After capturing this information the SD transmits these signals to the HSD. In the model presented in Figure \ref{fig:Fig_TactileModel_Generic}, the signals generated by the SD are expressed as
\begin{equation}
\mathbf{g}(n) =\left [g_1(n), \dots, g_i(n), \dots, g_{N_{SD}}(n)\right],
\end{equation}
where $g_i(n)$ is the $i$-th signal generated by the SD at the $n$-th instant of time, $t_s$ and $N_{SD}$ is the amount of signals. The HSD in turn processes this information and sends to the NW subsystem through the array $\mathbf{h}(n)$, expressed by
\begin{equation}
\mathbf{h}(n) = \left[h_1(n), \dots, h_i(n), \dots, h_{N^b_{HSD}}(n)\right],
\end{equation}
where $h_i(n)$ is the $i$-th signal generated by HSD at the $n$-th instant of time, $t_s$ and $N^b_{HSD}$ is the amount of signals.

The signal received by the HMD through the backwards channel of the NW subsystem can be expressed as
\begin{equation}
\mathbf{q}(n) = \left[q_1(n), \dots, q_i(n), \dots, q_{N^b_{HSD}}(n)\right],
\end{equation}
where
\begin{equation}
q_i(n) = h_i \left(n-d_i^b(n) \right) + r^b_i(n)
\end{equation}
in which, $r^b_i(n)$ represents an added noise and $d_i^b(n)$ represents a delay associated with the $i$-th information transmitted in $\mathbf{q}(n)$ by the backwards channel. Similarly to the forward channel, noise can also be characterized as a random variable Gaussian of zero mean and variance $\sigma^2_{rb}$ and delays are characterized as integers with $t_s$ granularity. The HMD processes the $\mathbf{q}(n)$ signal information and generates a set of control signals that will act on the MD and can be characterized as
\begin{equation}
\mathbf{p}(n) = \left[p_1(n), \dots, p_i(n), \dots, p_{N^b_{HMD}}(n)\right],
\end{equation}
where $p_i(n)$ is the $i$-th signal generated by the HMD at the $n$-th instant of time $t_s$ and $N^b_{HMD}$ is the number of signals. The MD in turn will synthesize the reaction stimuli generated by the environment, i.e. the ENV subsystem. Based on the model, it is possible to characterize these reaction stimuli as a signal expressed by
\begin{equation}
\mathbf{\hat{o}}(n) = \left[ \hat{o}_1(n), \dots, \hat{o}_i(n), \dots, \hat{o}_{N_{ENV}}(n) \right],
\end{equation}
where $\hat{o}_i(n)$ is an estimate of the $i$-th stimulus $o_i(n)$ generated in the ENV subsystem. Examples of reaction stimuli generated or synthesized by MD are touch, strength and temperature.

In addition to the latency associated with the NW subsystem that characterizes the communication medium between the OP and ENV subsystems, the MD, HMD, HSD, and SD subsystems also add latency to the system. Based on the work presented in \cite{tactile_network2015_TowardsLatency, internetSkills_mischa2017} these components represent $30\%$ of total latency. The latency of the MD and SD subsystems are associated with sensors and actuators that can be mechanical, electrical, electromechanical and other variations. HMD and HSD latencies are associated with the processing time of the algorithms in these devices and depending on the type of hardware and implementation architecture this latency can be considerably reduced.

\section{PHANToM Omni Device Model (MD \& SD)}\label{sec:ModelHapticDevice}

Based on the scheme presented in Figure \ref{fig:Fig_TactileModel_Generic}, this section presents details associated with the MD and SD used as reference for the hardware system proposed in this research. The MD and SD are characterized as a three degree of freedom robotic manipulator, 3-DoF, called the PHANToM Omni \cite{phantom_guide} (Figure \ref{fig:OmniPhantomEstrutura}). The PHANToM Omni has been widely used in literature as presented in \cite{2006_OmniPhantom_Teleoperation} and \cite{2012_OmniPhantom_Teleoperation}. In this work two of this devices are going to be used: one as an MD and the other as a SD.

\Figure[ht]()[width=0.9\linewidth]{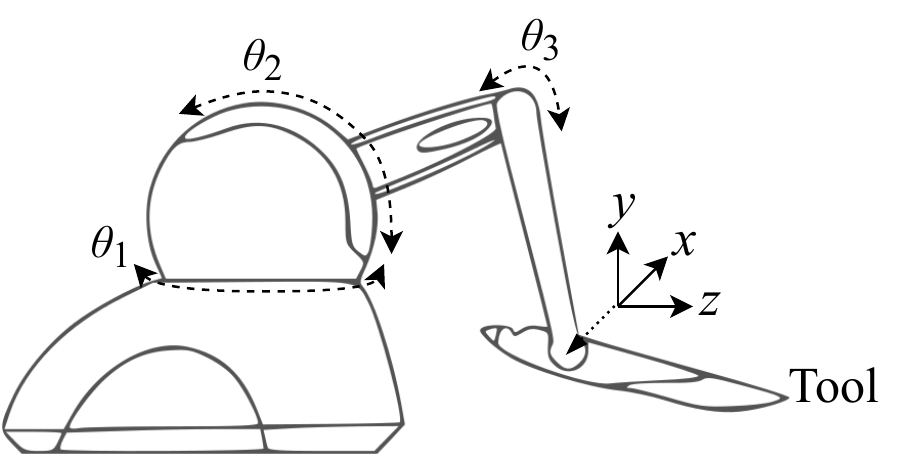}
{PHANToM Omni - MD and SD.\label{fig:OmniPhantomEstrutura}}

As can be seen from Figure \ref{fig:OmniPhantom}, the PHANToM Omni physical structure is formed by a base, an arm with two segments $L_1$ and $L_2$ which are interconnected by three rotary joints $\theta_1$, $\theta_2$ and $\theta_3$ and a tool. The variables presented in Figure \ref{fig:OmniPhantom} are represented by: $L_1$ = 0.135mm, $L_2$ = $L_1$, $L_3$=0.025mm and $L_4=L_1+A$ where $A$=0.035mm as described in \cite{phantom2009_Phantom}. These detailed features of the device are essential for performing the kinematics and dynamic calculations.

\Figure[ht]()[width=0.9\linewidth]{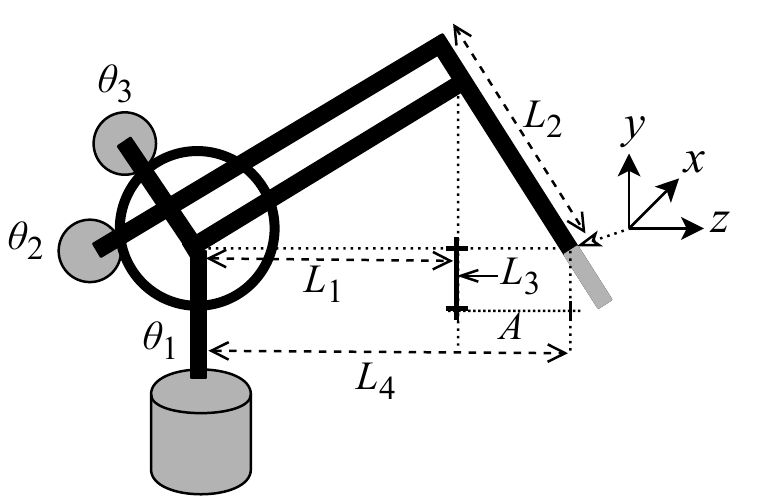}
{PHANToM Omni structure - MD and SD.\label{fig:OmniPhantom}}

\subsection{Forward Kinematics}\label{FKSection}

The kinematics of manipulative devices makes use of the relationship between operational coordinates and joint coordinates. Forward kinematics (FK) correlates the angular variables of the joints with the Cartesian system. That is, given an array of joint coordinates it is possible to determine the spatial position of the tool through the equation that can be expressed by
\begin{equation} \label{eq:X}
x = - \sin(\theta_1) (L_2\sin(\theta_3) + L_1 \cos(\theta_2)),
\end{equation}
\begin{equation} \label{eq:Y}
y = - L_2\cos(\theta_3)+ L_1\sin(\theta_2) + L_3,
\end{equation}
\begin{equation} \label{eq:Z}
\begin{aligned}
z = L_2\cos(\theta_1) \sin(\theta_3) \\ + L_1\cos(\theta_1)\cos(\theta_2) - L_4
\end{aligned}
\end{equation}
where $x$, $y$ and $z$ are variables that determine the spatial position of the tool in the Cartesian plane.

\subsection{Inverse Kinematics}\label{IKSection}

In the inverse kinematics (IK), the relationship between the joint angles and the Cartesian system is reversed, that is, given the spatial position of the tool it may be possible to determine the joint coordinates. The solution to this process is not as straightforward as in the direct kinematics. In direct kinematics, the position of the tool is determined solely by the displacements of the joints. In inverse kinematics, equations are composed of nonlinear calculations formed by trigonometric functions. Depending on the manipulator structure, multiple solutions may be possible for the same tool position, or there may be no solution for a particular set of tool positions. Based on the works \cite{phantom2001_Kinematics}, \cite{phantom2006_AStudy} and \cite{phantom2009_Phantom}, the value of $\theta_1$ can be defined through the equation expressed by
\begin{equation} \label{eq:Theta1}
\theta_1 = - \text{atan}2\left(x ,z + L_4 \right)
\end{equation}
where $x$ and $z$ represent coordinates in the Cartesian plane and $L_4$ corresponds to the size of the the arm segments, as shown in Figure \ref{fig:OmniPhantom}.

To calculate the other two joints $\theta_2$ and $\theta_3$ it is necessary to perform intermediate calculations. Thus, one can obtain $R$, $r$, $\beta$, $\gamma$ and $\alpha$ through the equations
\begin{equation} \label{eq:Rn}
R = \sqrt{ x^2 + (z+L_4)^2 },
\end{equation}
\begin{equation} \label{eq:rn}
r = \sqrt{x^2+ z + L_4)^2 + (y-L_3)^2  },
\end{equation}
\begin{equation} \label{eq:Gama}
\gamma = \text{acos}\biggl(\frac{L_1^2 - L_2^2 + r^2}{2 L_1 r} \biggr),
\end{equation}
\begin{equation} \label{eq:Beta}
\beta(n) = \text{atan}2(y - L_3 , R),
\end{equation}
and
\begin{equation} \label{eq:Alpha}
\alpha = \text{acos}\biggl(\frac{L_1^2 + L_2^2 - r^2}{2 L_1 L_2}\biggr).
\end{equation}

After performing the intermediate calculations it is possible to calculate $\theta_2$ through the equation 
\begin{equation} \label{eq:Theta2}
\theta_2 = \gamma + \beta.
\end{equation}
Finally, the value corresponding to the $\theta_3$ joint can be obtained through the equation 
\begin{equation} \label{eq:Theta3}
\theta_3 = \theta_2 +\alpha - \frac{\pi}{2}.
\end{equation}

\subsection{Kinesthetic Feedback Force}\label{SecKFF}

The kinesthetic feedback force allows the environment to be "felt", i.e. when the SD comes into physical contact with an object, the MD will receive a counter force. This model can be implemented through the equation 
\begin{equation} \label{eq:KinestheticFeedbackForce}
\boldsymbol \tau = \mathbf{J}^T\mathbf{F},
\end{equation}
where $\boldsymbol \tau$ defines the torque array that will be applied to each joint ($\theta_1$, $\theta_2$ and $\theta_3$) of the PHANToM Omni associated with the MD, $\mathbf{J}^T$ is the transpose of the Jacobian matrix and $\mathbf{F}$ is the force array resulting from the interaction of SD with ENV. The torque array $\boldsymbol \tau$ can be expressed as
\begin{equation}
\boldsymbol \tau = \left[\tau_1, \tau_2, \tau_3\right].
\end{equation}

The $ \mathbf{J}$ Jacobian matrix incorporates structural information about the handler and it is identified as
\begin{equation} \label{eq:Jacobian}
\mathbf{J}
=
\begin{bmatrix}
J_{11} & J_{12} & J_{13} \\ 
J_{21} & J_{22} & J_{23} \\ 
J_{31} & J_{32}& J_{33}
\end{bmatrix},
\end{equation}
where 
\begin{equation} \label{eq:JM_11}
J_{11} = -\cos(\theta_1) ( L_2  \sin(\theta_3) + L_1 \cos(\theta_2) ),
\end{equation}
\begin{equation} \label{eq:JM_21}
J_{21} = 0,
\end{equation}
\begin{equation} \label{eq:JM_31}
J_{31} = -L_1 \cos(\theta_2) \sin(\theta_1) - L_2 \sin(\theta_3)  \sin(\theta_1),
\end{equation}
\begin{equation} \label{eq:JM_12}
J_{12} = L_1 \sin(\theta_1) \sin(\theta_2),
\end{equation}
\begin{equation} \label{eq:JM_22}
J_{22} = L_1 \cos(\theta_2),
\end{equation}
\begin{equation} \label{eq:JM_32}
J_{32} = - L_1 \sin(\theta_2) \cos(\theta_1),
\end{equation}
\begin{equation} \label{eq:JM_13}
J_{13} = - L_2 \sin(\theta_1) \cos(\theta_3),
\end{equation}
\begin{equation} \label{eq:JM_23}
J_{23} = L_2 \sin(\theta_3),
\end{equation}
and
\begin{equation} \label{eq:JM_33}
J_{33} = L_2 \cos(\theta_3) \cos(\theta_1).
\end{equation}

The force array $\mathbf{F}$ is expressed as
\begin{equation} \label{eq:Force}
\mathbf{F} = \left[F_x, F_y, F_z\right]
\end{equation}
and can be obtained through sensors internal or external to the device. According to (\ref{eq:KinestheticFeedbackForce}), the $\boldsymbol \tau$ torque array representing the resulting force at each joint can be defined as
\begin{equation} \label{eq:Tau_1}
\tau_1 = J_{11} F_x + J_{21} F_y + J_{31} F_z,
\end{equation}
\begin{equation} \label{eq:Tau_2}
\tau_2 = J_{12} F_x + J_{22} F_y + J_{32}F_z,
\end{equation}
and
\begin{equation} \label{eq:Tau_3}
\tau_3= J_{13} F_x + J_{23} F_y + J_{33} F_z.
\end{equation}

\section{Simulated Tactile Internet Model}\label{sec:SimulatedTactileModel}

Figures \ref{fig:Fig_TactileModel_Generic} and \ref{fig:Fig_TactileModel_Detailed} details the structure used for the hardware design in FPGA, in which a given operator, OP, handles a  PHANToM Omni on the master side, MD, which is connected to HMD that, in this case, is a dedicated FPGA hardware. Data is transmitted through the network, the NW subsystem, to HSD which is also a dedicated hardware in FPGA. The HSD is also connected to a PHANToM Omni that interacts with the environment, the ENV subsystem. Figure \ref{fig:Fig_TactileModel_Detailed} also details the backwards direction from the ENV and the OP.

\Figure[ht]()[width=\linewidth]{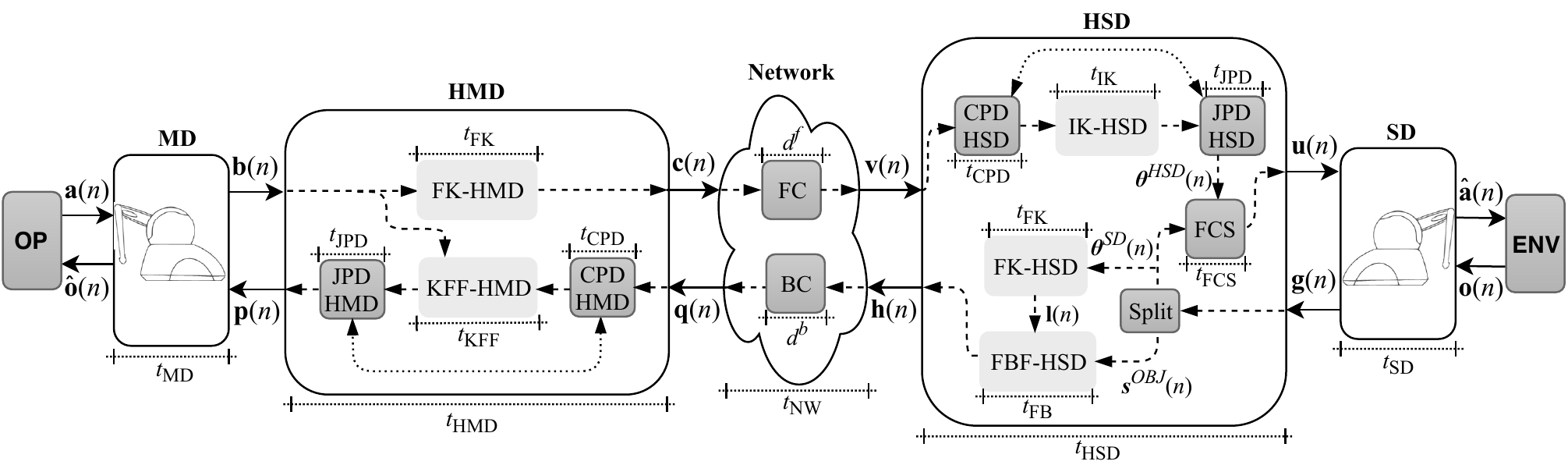}
{Detailed discrete model of a tactile internet system.\label{fig:Fig_TactileModel_Detailed}}

The OP is modeled as an information source responsible for generating a spatial trajectory through discrete signals expressed in the $\mathbf{a}(n)$ array. At each $n$-th instant $t_s$ the OP sends three variables $x^{OP}(n)$, $y^{OP}(n)$ and $z^{OP}(n)$ representing the positioning of the MD tool (Figures \ref{fig:OmniPhantomEstrutura} and \ref{fig:OmniPhantom}) in the Cartesian space an this is expressed by
\begin{equation} \label{eq:Signal_A}
\mathbf{a}(n) =\left[x^{OP}(n), y^{OP}(n), z^{OP}(n)\right].
\end{equation}
This step simulates the spatial movement of the MD tool by the operator, that is, at each instant of time, $t_s$, a spatial movement is performed and a new signal $\mathbf{a}(n)$ is generated by the OP.

The PHANToM Omni has encoders at its three joints that translate spatial positioning at the three angles $\theta_1$, $\theta_2$ and $\theta_3$ ( Figures \ref{fig:OmniPhantomEstrutura} and \ref{fig:OmniPhantom}). Thus, based on Figure \ref{fig:Fig_TactileModel_Detailed}, it can be said that MD converts the signal $\mathbf{a}(n)$ into a signal expressed as
\begin{equation} \label{eq:Signal_B}
\mathbf{b}(n) = \left[\theta^{MD}_1(n), \theta^{MD}_2(n), \theta^{MD}_3(n)\right]
\end{equation}
and forwards it to the HMD at every $n$-th instant of time $t_s$. 

Then, as can be seen in Figure \ref{fig:Fig_TactileModel_Detailed}, the $\mathbf{b}(n)$ signal propagates to the HMD, which on receiving the signal transforms the joint positioning angles, $\mathbf{b}(n)$, into spatial position by calculating the FK according to (\ref{eq:X}), (\ref{eq:Y}) and (\ref{eq:Z}). All equations are implemented in FPGA through a hardware module called the FK-HMD. The equations are implemented in parallel which can significantly increase the processing time. The use of FK is motivated by an reduction of the amount of information utilized, i.e., for a $N$-DoF robotic manipulator $N$ joint angles will be generated and that can be converted into only three values associated with the spatial position of the tool,  $x$, $y$ and $z$. On the other hand, the use of this strategy increases the amount of calculations to be performed by the MD, which is compensated by the parallel implementation of the algorithm in FPGA. It is essential to note that the use of custom hardware operating in parallel allows processing time not to be substantially affected by $N$.

Based on Section \ref{sec:GenericTactileModel}, after the FK calculation by the FK-HMD hardware module, a new discrete signal is created that can be expressed by
\begin{equation} \label{eq:Signal_C}
\mathbf{c}(n) = \left[x^{HMD}(n), y^{HMD}(n), z^{HMD}(n)\right]
\end{equation}
where $x^{HMD}(n)$, $y^{HMD}(n)$ and $z^{HMD}(n)$ are the values of the spatial coordinate array generated by the HMD to be sent to HSD via the communication medium, NW. The FK-HMD hardware module generates a new $\mathbf{c}(n)$ array every $n$-th instant of time.

After the transmission through the forward channel, here called FC, the signal received by the HSD can be expressed as
\begin{equation} \label{eq:Signal_V}
\mathbf{v}(n) = \left[x^{HSD}(n), y^{HSD}(n), z^{HSD}(n)\right].
\end{equation}
Based on (\ref{EQChannel1}) the spatial coordinate signal received by HSD can be expressed as
\begin{equation}\label{canalFC1}
x^{HSD}(n) = x^{HMD} \left(n-d^f_x(n) \right) + r^{f}_x(n),
\end{equation}
\begin{equation}\label{canalFC2}
y^{HSD}(n) = y^{HMD} \left(n-d^f_y(n) \right) + r^{f}_y(n),
\end{equation}
and
\begin{equation}\label{canalFC3}
z^{HSD}(n) = z^{HMD} \left(n-d^f_z(n) \right) + r^{f}_z(n)
\end{equation}
where $d^f_x(n)$, $d^f_y(n)$, $d^f_z(n)$, $r^{f}_x(n)$, $r^{f}_y(n)$ and $r^{f}_z(n)$ are the delays and noises associated with CF.

As in this case the Slave PHANToM Omni, SD, copies the movement of the master PHANToM Omni, MD, it is necessary for the HSD to perform a feedback control system on the three joints of the PHANToM Omni slave, here expressed as
\begin{equation} \label{eq:Signal_Theta_SD}
\boldsymbol \theta^{SD}(n) = \left[\theta_1^{SD}(n), \theta_2^{SD}(n), \theta_3^{SD}(n)\right]
\end{equation}
that is, $\theta_1^{SD}(n)$, $\theta_2^{SD}(n)$, $\theta_3^{SD}(n)$
are control variables associated with DS. The control system illustrated in Figure \ref{fig:Fig_TactileModel_Detailed} as FCS shall minimize the error, $\mathbf{e}^{FCS}(n)$, between $\boldsymbol \theta^{SD}(n)$ and the reference signal $\boldsymbol \theta^{HSD}(n)$ characterized as
\begin{equation} 
\boldsymbol \theta^{HSD}(n) = \left[\theta_1^{HSD}(n), \theta_2^{HSD}(n), \theta_3^{HSD}(n)\right]
\end{equation}
where 
\begin{equation} \label{eq:Signal_E}
\mathbf{e}(n) = \boldsymbol \theta^{HSD}(n) - \boldsymbol \theta^{SD}(n) and
\end{equation}
and
\begin{equation}
\left[
\begin{array}{c}
e^{FCS}_1(n) \\
e^{FCS}_2(n) \\
e^{FCS}_3(n)
\end{array} \right] = \left[
\begin{array}{c}
\theta_1^{HSD}(n) \\
\theta_2^{HSD}(n) \\
\theta_3^{HSD}(n) 
\end{array} \right] - \left[
\begin{array}{c}
\theta_1^{SD}(n) \\
\theta_2^{SD}(n) \\
\theta_3^{SD}(n) 
\end{array} \right].
\end{equation}

The $\boldsymbol \theta^{SD}(n)$ signal is obtained from the SD via sensors (encoders) at the SD joints and the $\boldsymbol \theta^{HSD}(n)$ signal is obtained from the IK-HSD hardware module shown in Figure \ref{fig:Fig_TactileModel_Detailed}. This hardware module implements all inverse kinematics equations presented in Section \ref{IKSection}, i.e. (\ref{eq:Theta1}) through (\ref{eq:Theta3}). There are several techniques and approaches that can be used in the FCS module ranging from more traditional techniques such as a proportional–integral–derivative controller \cite{8343385} to more innovative artificial intelligence based techniques \cite{Yang2016, Nazemizadeh2014}.

The CPD-HSD and JPD-HSD modules, illustrated in Figure \ref{fig:Fig_TactileModel_Detailed}, represent the algorithms of prediction and detection in cartesian space and joints, respectively. These modules are responsible for minimizing the latency and noise added by the FC associated with the tactile internet system (Eqs. (\ref{canalFC1}), (\ref{canalFC2}) and (\ref{canalFC3})). Depending on the prediction and detection technique used, the HSD may use only one of the modules, namely the CPD-HSD or JPD-HSD. There is still no consensus about whether the Cartesian space or joints is the best for minimizing latency and noise inserted by the channel. There are several works in the literature that present proposals using only one of the spaces and proposals that try to use the information from both simultaneously. 

Similarly to the FCS module, approaches ranging from the more traditional techniques up to more innovative techniques based on artificial intelligence have been used in the CPD-HSD and JPD-HSD modules \cite{SaiHong2014, app8122648, Xiang2019, 6094666, electronics8040398}. Thus, it can be said that $\boldsymbol \theta^{HSD}(n)$ is an estimate of the $\mathbf{b}(n)$ signal generated by the MD.

At each $n$-th time, the FCS acts on the SD through the $\mathbf{u}(n)$ signal, detailed in Figures \ref{fig:Fig_TactileModel_Generic} and \ref{fig:Fig_TactileModel_Detailed}, which in the case of the PHANToM Omni can be expressed as
\begin{equation} 
\mathbf{u}^{HSD}(n) = \left[\tau_1^{HSD}(n), \tau_2^{HSD}(n), \tau_3^{HSD}(n)\right]
\end{equation}
where $\tau_i^{HSD}(n)$ is the $i$-th torque applied every $i$-th joint. The FCS will act as a tracking mechanism, making the SD follow the path traveled by the MD. Finalizing the data stream associated with the forward channel, it can be said that the $\mathbf{\hat{a}}(n)$ signal is formed by an estimate of the spatial position generated by the OP, $\mathbf{\hat{a}}(n)$, i.e.
\begin{equation} 
\mathbf{\hat{a}}(n) =\left[\hat{x}^{OP}(n), \hat{y}^{OP}(n), \hat{z}^{OP}(n)\right].
\end{equation}

The interaction of the PHANToM Omni, SD, with ENV can vary from free movement to physical contact. When some kind of physical contact occurs, the SD detects the touch and sends this information back to the HSD. As per the model detailed in Figure \ref{fig:Fig_TactileModel_Detailed} the ENV sends back to SD the information associated with the contact force in the spatial plane, expressed here as,
\begin{equation} 
\mathbf{o}(n) =\left[F_x^{ENV}(n), F^{ENV}_y(n), F^{ENV}_z(n) \right].
\end{equation}
The value associated with the contact force information can be measured directly through SD-coupled force sensors or indirectly estimated through other types of sensors that may be SD-coupled or inserted into the environment \cite{FeedbackForce2018}. In the case of the model presented in Figure \ref{fig:Fig_TactileModel_Detailed}, the SD sends to HSD the objects surface's spatial positions through sensors spread in the ENV. The signal expressed as
\begin{equation} 
\mathbf{s}^{OBJ}(n) =\left[x^{OBJ}(n), y^{OBJ}(n), z^{OBJ}(n) \right]
\end{equation}
represents the spatial position of the closest object from the SD tool. Thus, based on the information already described, every $n$-th time $t_s$ the SD sends to the HSD a signal characterized by the array $\mathbf{g}(n)$ expressed as
\begin{equation} \label{eq:Signal_G}
\mathbf{g}(n) =\left[\boldsymbol \theta^{SD}(n), \mathbf{s}^{OBJ}(n)\right].
\end{equation}

In the HSD, when the signal $\mathbf{g}(n)$ is received, the Split module separates the $\boldsymbol \theta^{SD}(n)$ signal and sends it to the FCS and the FK-HSD hardware module. And the signal $\mathbf{s}^{OBJ}(n)$ is sent to the FB-HSD hardware module, as detailed in Figure \ref{fig:Fig_TactileModel_Detailed}. The FK-HSD hardware module performs the forward kinematics calculation similarly to FK-HMD and thus the current spatial position of the SD tool in the environment, ENV, can be obtained. Every $n$-th instant $t_s$ FK-HSD generates a signal expressed as
\begin{equation} \label{eq:Signal_I}
\mathbf{l}(n) = [x^{ENV}(n), y^{ENV}(n), z^{ENV}(n)]
\end{equation}
where $x^{ENV}(n)$, $y^{ENV}(n)$ and $z^{ENV}(n)$
are the spatial position of the tool in the ENV module from $\boldsymbol \theta^{SD}(n)$. The FBF-HSD hardware module implements the calculations associated with the generation of the feedback force from the contact between the tool and the object. Based on the work presented in \cite{FeedbackForce2018} the contact force, represented by the $\mathbf{h}(n)$ signal, can be expressed as
\begin{equation} \label{eq:ControledeForca}
\mathbf{h}(n) =\left [F^{HSD}_x(n), F^{HSD}_y(n), F^{HSD}_z(n)\right],
\end{equation}
where
\begin{equation} \label{eq:FF_Fx}
F^{HSD}_x(n) =  h_x(n)\left( x^{OBJ}(n) - x^{ENV}(n)\right),
\end{equation}
\begin{equation} \label{eq:FF_Fy}
F^{HSD}_y(n) =  h_y(n)\left( y^{OBJ}(n) - y^{ENV}(n)\right),
\end{equation}
and
\begin{equation} \label{eq:FF_Fz}
F^{HSD}_z(n) =  h_z(n)\left( z^{OBJ}(n) - z^{ENV}(n)\right).
\end{equation}
In these equations, the constants $h_x(n)$, $h_y(n)$ and $h_z(n)$ represent the elasticity coefficients associated with the object. It is important to note that in this model the $\mathbf{h}(n)$ signal is a synthesized version of the real force value here characterized by the $\mathbf{o}(n)$ array.

After the feedback force calculation process, as illustrated in Figure \ref{fig:Fig_TactileModel_Detailed}, the $\mathbf{h}(n)$ signal is transmitted to the HMD via the backwards channel (BC) which, similarly to FC, adds latency and noise. The signal received by the HMD can be expressed as
\begin{equation} 
\mathbf{q}(n) = [F^{HMD}_x(n), F^{HMD}_y(n), F^{HMD}_z(n)]
\end{equation}
where 
\begin{equation}\label{canalBC1}
F^{HMD}_x(n) = F_x^{HSD} \left(n-d^b_x(n) \right) + r^{b}_x(n),
\end{equation}
\begin{equation}\label{canalBC2}
F_y^{HMD}(n) = F_y^{HSD} \left(n-d^b_y(n) \right) + r^{b}_y(n),
\end{equation}
and
\begin{equation}\label{canalBC3}
F_z^{HMD}(n) = F_z^{HSD} \left(n-d^b_z(n) \right) + r^{b}_z(n)
\end{equation}
where $d^b_x(n)$, $d^b_y(n)$, $d^b_z(n)$, $r^{b}_x(n)$, $r^{b}_y(n)$ and $r^{b}_z(n)$ are the latencies and the noises associated with the BC.

Similarly to HSD, the HMD will minimize the effect of latency and noise from operations of Cartesian and joint space. For HMD, the calculations associated with the Cartesian space will be performed by the CPD-HMD module and associated with the joint space by the JPD-HMD module. In addition to the prediction and detection calculations, the HMD must transform the force signals received through signal $\mathbf{q}(n)$ into a torque to be applied to the MD joints which is accomplished by the KFF-HMD hardware module. KFF-HMD implements the equations (\ref{eq:Tau_1}), (\ref{eq:Tau_2}) and (\ref{eq:Tau_3}) presented in Section \ref{SecKFF} and generate the signal expressed as
\begin{equation} 
\mathbf{p}(n) = \left[\tau_1^{HMD}(n), \tau_2^{HMD}(n), \tau_3^{HMD}(n)\right]
\end{equation}
where $\tau_i^{HMD}(n)$ is the torque associated with the $i$-th joint of the MD. Since the PHANToM Omni is a haptic device, it already has a built-in control system, FCS, which uses as reference signal the torques associated with the $\mathbf{p}(n)$ array.

After applying the torques to the MD joints via the $\mathbf{p}(n)$ signal, the OP receives the feedback force signal, in other words, it feels the object touched by the SD in the ENV. This sensation is identified in by the $\mathbf{\hat{o}}(n)$ signal expressed as
\begin{equation} 
\mathbf{\hat{o}}(n) = \left[\hat{F}_x^{ENV}(n), \hat{F}^{ENV}_y(n), \hat{F}^{ENV}_z(n) \right].
\end{equation}

As illustrated in Figure \ref{fig:Fig_TactileModel_Detailed}, the MD, HMD, NW, HSD, and SD subsystems have the following runtimes: $t_{MD}$, $t_{HMD}$, $t_{NW}$, $t_{HSD}$ and $t_{SD}$, respectively. The sum of these, times taking into account the forward direction (between OP and ENV) and the backwards direction (between ENV and OP), represent the total system latency that can be expressed as
\begin{equation}
t_{\text{latency}} = 2\left(t_{\text{MD}} + t_{\text{HMD}} + t_{\text{NW}} + t_{\text{HSD}} + t_{\text{SD}}\right).
\end{equation}
Some works presented in the literature review agree that the ideal requirement is that $t_{\text{latency}} \leq  1 \, \text{ms}$, on the other hand, other works point out that the latency requirement can be expresses as $t_{\text{latency}} \leq 10 \, \text{ms}$, depending on the application \cite{REF_TIME_MS_01, 8399482, REF_TIME_MS_02, tactile_network2016_5GEnable, s19225029}. Considering that $30\%$ of the total latency time $t_{\text{latency}}$ is spent by MD, HMD, HSD, and SD, it can be understood that
\begin{equation}
\left(t_{\text{MD}} + t_{\text{HMD}} + t_{\text{HSD}} + t_{\text{SD}}\right) \leq \frac{0.3t_{\text{latency}}}{2}.
\end{equation}
Assuming an equal time division among MD, HMD, HSD, and SD it is possible to affirm that the time associated with hardware, $t_{\text{hardware}}$, whether the master, HMD, or the slave device, HSD, can be expressed as
\begin{equation}\label{timeHardware}
t_{\text{HMD}} = t_{\text{HSD}} = t_{\text{hardware}}  \leq \frac{0.3t_{\text{latency}}}{8}.
\end{equation}
Taking the $1 \, \text{ms}$ constraints into consideration and substituting this value in (\ref{timeHardware}), it is possible to affirm that the hardware time, $t_{\text{hardware}}$, must meet the  $t_{\text{hardware}} \leq 37.5 \, \mu\text{s}$ constraint for all cases (condition $1 \, \text{ms}$) or the $t_{\text{hardware}} \leq 375 \, \mu\text{s}$ constraint for some specific cases ( $10 \, \text{ms}$ condition).

Recent studies from the literature show that the $1 \, \text{ms}$ restriction ($t_{\text{hardware}} \leq 37.5 \, \mu\text{s}$) is  difficult to achieve using hardware devices based on embedded systems such as microprocessors and microcontrollers \cite{REF20_RelatedWork04, REF19_RelatedWork03}. The $10 \, \text{ms}$ restriction ($t_{\text{hardware}} \leq 375 \, \mu\text{s}$) is achieved in specific cases where SD is a virtual environment and HSD is a high performance processor computer \cite{s19225029}. Thus this work aims to minimize the execution time in HMD, $t_{\text{HMD}}$, and HSD, $t_{\text{HSD}}$, using FPGA reconfigurable computation. In other words, the target is to achieve a $t_{\text{hardware}} \leq 37.5 \, \mu\text{s}$.

This paper presents a hardware reference model for the FK-HMD, KFF-HMD, IK-HSD, FK-HSD, and FBF-HSD modules illustrated in Figure \ref{fig:Fig_TactileModel_Detailed}. The complete model that will be presented in detail in the next section makes use of a parallel implementation methodology in which high throughput is prioritized, i.e. the execution time of the modules $t_{\text{FK}}$, $t_{\text{KFF}}$, $t_{\text{IK}}$ and $t_{\text{FBF}}$, illustrated in Figure \ref{fig:Fig_TactileModel_Detailed}.

This work does not propose dedicated hardware reference models for the CPD-HSD, JPD-HSD, CPD-HMD, JPD-HMD and FCS modules as there are several techniques and algorithms that can be applied to them. However, considering the hardware time constraints, $t_{\text{hardware}}$, it is noted that it is also important to use dedicated hardware structures with reconfigurable computing for these modules. Studies in the literature foresee the use of AI based techniques for these modules; however, it is essential to note that AI techniques and algorithms implemented on general purpose processor-based hardware platforms can lead to higher processing times \cite{marceloAlisson2014, 8626462, 8678408, Torquato2019, 8574886, electronics8060631, NORONHA2019138}.

\section{Implementation Description} 
\label{sec:Implementation}

The FK-HMD and KFF-HMD hardware modules associated with the master device (HMD) and the IK-HSD, FK-HSD, and FBF-HSD hardware modules associated with the slave device (HSD) (Figure \ref{fig:Fig_TactileModel_Detailed}) were designed using a parallel implementation in order to prioritize the processing speed. The implementations were designed in FPGA using a hybrid scheme with fixed point and floating point representation in distinct parts of the proposed architecture. In the portions that adopt the fixed point format, the variables follow a notation expressed as $[sV.N]$ indicating that the variable is formed by $V$ bits of which $N$ bits are intended for the fractional part and the $s$ symbol indicates that the variable is signed. In this case, the number of bits intended for the integer part is $V-N-1$. For the representation of floating point variables, the notation [F32] is adopted. Most of the implemented circuits were designed using a $32$-bit single precision (IEEE754) floating point format representation. The fixed point format was used only on the circuit that implements the trigonometric function block (TFB) module, as illustrated in Figure \ref{fig:Fig_IMP_CORDIC}. TFB is the module responsible for performing trigonometric operations through the hardware implementation of CORDIC (\emph{COordinate Rotation DIgital Computer}) \cite{algoritmoCordic}. The implemented CORDIC circuit uses data representation in fixed point format using the $[s16.13]$ representation.

\Figure[ht]()[width=0.9\linewidth]{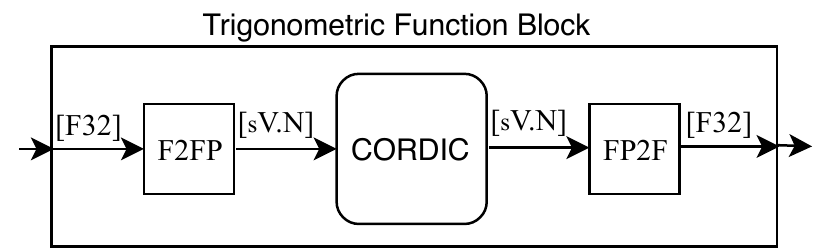}
{Proposed circuit for calculating trigonometric functions - TFB.\label{fig:Fig_IMP_CORDIC}}

As illustrated in Figure \ref{fig:Fig_IMP_CORDIC}, the TFB module receives data from external circuits in the $32$-bits floating point standard. A conversion to the fixed point numeric representation type represented by the $[s16.13]$ notation is performed through the Float to Fixed-point (F2FP) module that has been implemented in hardware. After the CORDIC hardware operations are performed, the data in the fixed point format is transformed back to the $32$-bit floating point through the Fixed-point to Float (FP2F) module which was also implemented in hardware.

Several of the proposed methods to be presented use the constants $L_1$, $L_2$, $L_3$ and $L_4$. They represent physical characteristics of the PHANToM Omni device as illustrated in Figure \ref{fig:OmniPhantomEstrutura}. These constants use the $32$-bit floating point numeric representation.

\subsection{FORWARD KINEMATICS (FK-HMD and FK-HSD))}
\label{sec_FK-HMD}

As illustrated in Figure \ref{fig:Fig_TactileModel_Detailed}, both the hardware associated with the master device (HMD) and the hardware associated with the slave device (HSD) implement forward kinematics through the FK-HMD and FK-HSD modules, respectively. These modules have the same FPGA-implemented circuit, differing only in the input and output signals. They are designed to work with three input signals, one for each component of the angular positioning of the device's joints, and three output signals, one for each component of the the positioning of the device's tool in the Cartesian system. The input signals are $\theta_1[F32](n)$, $\theta_2[F32](n)$ and $\theta_3[F32](n)$ and the output signals are $x[F32](n)$, $y[F32](n)$ and $z[F32](n)$. For FK-HMD, the input signals represent the $\theta_1^{MD}[F32](n)$, $\theta_2^{MD}[F32](n)$ and $\theta_3^{MD}[F32](n)$ signals, and the output signals represent the $x^{HMD}[F32](n)$, $y^{HMD}[F32](n)$ and $z^{HMD}[F32](n)$ signals. In the case of the FK-HSD module, the input signals represent the signals $\theta_1^{SD}[F32](n)$, $\theta_2^{SD}[F32](n)$ and  $\theta_3^{SD}[F32](n)$ and the output signals represent the signals $x^{ENV}(n)$, $y^{ENV}(n)$ and $z^{ENV}(n)$. At every $n$-th instant all the computation performed in order to calculate the forward kinematics are executed in parallel.

Based on (\ref{eq:X}), the algorithm used for calculating $x[F32](n)$ was implemented in FPGA through the generic circuit illustrated in Figure \ref{fig:Fig_IMP_CD_X}. The circuit was designed to work with three input signals $\theta_1[F32](n)$, $\theta_2[F32](n)$ and $\theta_3[F32](n)$ and one output signal. These signals are forwarded to TFB sub circuits where sine and cosine calculations are performed. For this process the constants $L_1$ and $L_2$, three multipliers, one inverter and one adder are used.

\Figure[ht]()[width=0.9\linewidth]{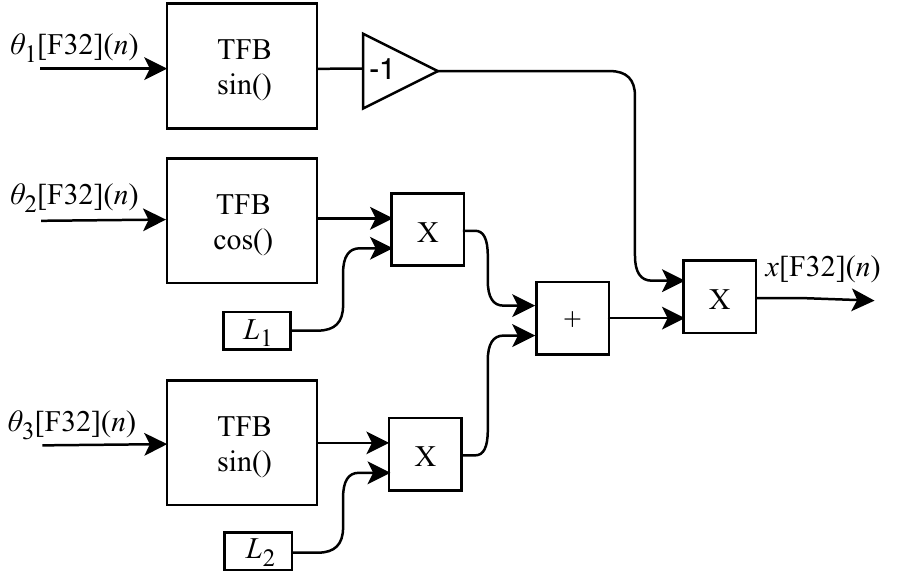}
{Proposed forward kinematics circuit for obtaining the $x[F32](n)$ spatial coordinate (Eq. (\ref{eq:X})) - FK-HMD and FK-HSD.\label{fig:Fig_IMP_CD_X}}

The calculation of $y[F32](n)$ based on (\ref{eq:Y}) was implemented in FPGA through the generic circuit shown in Figure \ref{fig:Fig_IMP_CD_Y}. The circuit was designed to work with two input signals $\theta_2[F32](n)$ and $\theta_3[F32](n)$ and one output signal. These signals are routed to TFB sub circuits to perform sine and cosine calculations. In the process flow two multipliers, two adders, one inverter and the constants $L_1$ and $L_2$ are used.

\Figure[ht]()[width=0.9\linewidth]{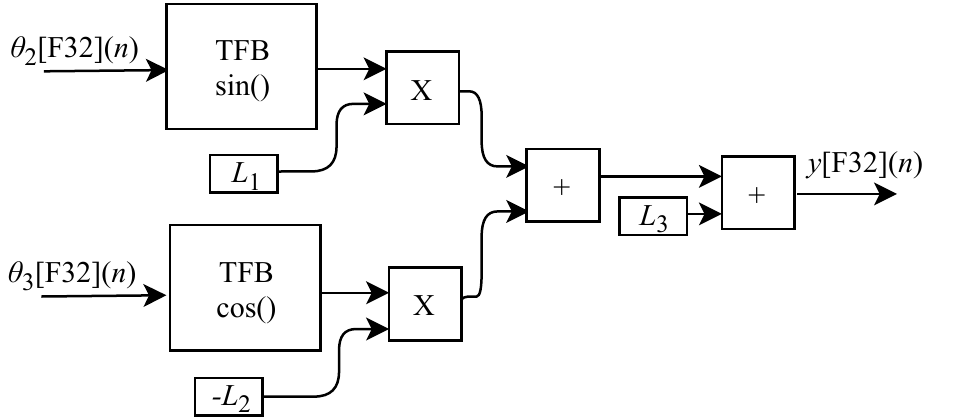}
{Proposed forward kinematics circuit for obtaining the $y[F32](n)$ spatial coordinate (Eq. (\ref{eq:Y})) - FK-HMD and FK-HSD.\label{fig:Fig_IMP_CD_Y}}

The generic circuit illustrated in Figure \ref{fig:Fig_IMP_CD_Z} was implemented in FPGA to perform the calculation of $z[F32](n)$ and it is based on (\ref{eq:Z}). The circuit is designed to work with three input signals $\theta_1[F32](n)$, $\theta_2[F32](n)$ and $\theta_3[F32](n)$ and one output signal. These signals are routed to TFB sub circuits in order to perform sine and cosine calculations. In the process flow four multipliers, two adders, one inverter and the constants $L_1$, $L_2$ and $L_4$ are used.

\Figure[ht]()[width=0.9\linewidth]{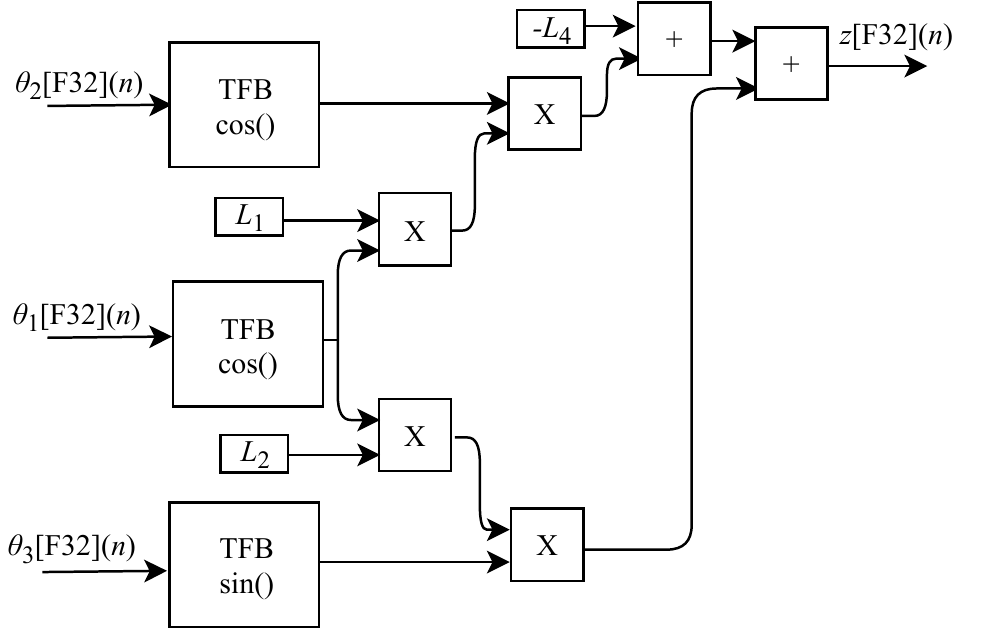}
{Proposed forward kinematics circuit for obtaining the $z[F32](n)$ spatial coordinate (Eq. (\ref{eq:Z})) - FK-HMD and FK-HSD.\label{fig:Fig_IMP_CD_Z}}

In the FK-HMD module the $\theta_1^{MD}[F32](n)$, $\theta_2^{MD}[F32](n)$ and $\theta_3^{MD}[F32](n)$ input signals are received through the $\mathbf{b}(n)$ array ((\ref{eq:Signal_B}) in section \ref{sec:SimulatedTactileModel}), then all calculation are performed in parallel resulting in the $\mathbf{c}(n)$ array ((\ref{eq:Signal_C})  in section \ref{sec:SimulatedTactileModel}) with the $x^{HMD}[F32](n)$, $y^{HMD}[F32](n)$ and $z^{HMD}[F32](n)$ signals as shown in Figure \ref{fig:Fig_TactileModel_Detailed}. For the FK-HSD module the $\theta_1^{SD}[F32](n)$, $\theta_2^{SD}[F32](n)$ and  $\theta_3^{SD}[F32](n)$ input signals enter the module via the $\boldsymbol \theta^{SD}(n)$ array ((\ref{eq:Signal_Theta_SD}) in section \ref{sec:SimulatedTactileModel}) and after performing all parallel computations, the resulting signals $x^{ENV}(n)$, $y^{ENV}(n)$ and $z^{ENV}(n)$ are output from the module via the $\mathbf{l}(n)$ array ((\ref{eq:Signal_Theta_SD}) in section \ref{sec:SimulatedTactileModel}).

\subsection{INVERSE KINEMATICS (IK-HSD)}

The hardware associated with the slave device (HSD) implements the inverse kinematics through the IK-HSD module, as shown in Figure \ref{fig:Fig_TactileModel_Detailed}. The IK-HSD FPGA-implemented circuit is designed to work with three input signals $x^{HSD}[F32](n)$, $y^{HSD}[F32](n)$ and $z^{HSD}[F32](n)$ and three output signals $\theta_1^{HSD}[F32](n)$, $\theta_2^{HSD}[F32](n)$ and $\theta_3^{HSD}[F32](n)$. However, to calculate $\theta_2^{HSD}[F32](n)$ (Eq. (\ref{eq:Theta2})) and $\theta_3^{HSD}[F32](n)$ (Eq. (\ref{eq:Theta3})) it is first necessary to perform intermediate calculations to obtain the values of $R[F32](n)$, $r[F32](n)$, $\beta[F32](n)$, $\gamma[F32](n)$ and $\alpha[F32](n)$

Based on (\ref{eq:Theta1}), (\ref{eq:Theta2}) and (\ref{eq:Theta3}), algorithms for calculating $\theta_1^{HSD}[F32](n)$, $\theta_2^{HSD}[F32](n)$ and $\theta_3^{HSD}[F32](n)$ were implemented in FPGA through the generic circuits illustrated in Figures \ref{fig:Fig_IMP_CI_Theta1}, \ref{fig:Fig_IMP_CI_Theta2} and \ref{fig:Fig_IMP_CI_Theta3} respectively.

\Figure[ht]()[width=0.8\linewidth]{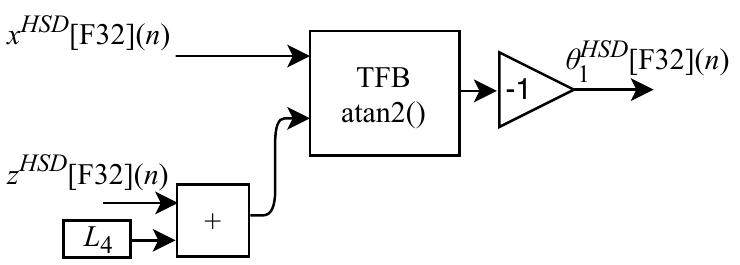}
{Proposed inverse kinematics circuit for obtaining the $\theta_1^{HSD}[F32](n)$ angular position (Eq. (\ref{eq:Theta1})) - IK-HSD.\label{fig:Fig_IMP_CI_Theta1}}

\Figure[ht]()[width=0.50\linewidth]{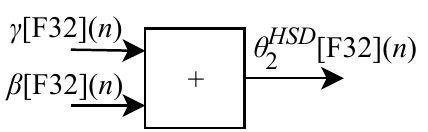}
{Proposed inverse kinematics circuit for obtaining the $\theta_2^{HSD}[F32](n)$ angular position (Eq. (\ref{eq:Theta2})) - IK-HSD.\label{fig:Fig_IMP_CI_Theta2}}

\Figure[ht]()[width=0.85\linewidth]{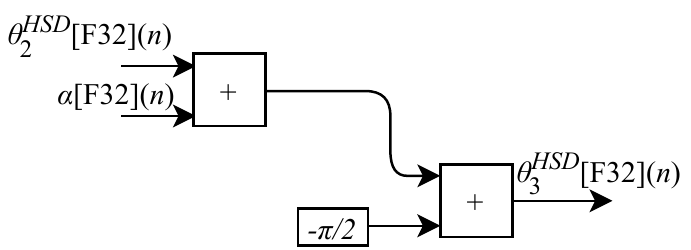}
{Circuito proposto da cinemática inversa para obter a posição angular $\theta_3^{HSD}[F32](n)$ (Eq. (\ref{eq:Theta3})) - IK-HSD.\label{fig:Fig_IMP_CI_Theta3}}

As already described, and according to the illustrations shown in Figures \ref{fig:Fig_IMP_CI_Theta2} and \ref{fig:Fig_IMP_CI_Theta3}, to perform the calculations of $\theta_2^{HSD}[F32](n)$ and $\theta_3^{HSD}[F32](n)$ it is first necessary to perform the intermediate calculations of $\gamma[F32](n)$ (Eq. (\ref{eq:Gama})), $\beta[F32](n)$ (Eq. (\ref{eq:Beta})) and $\alpha[F32](n)$ (Eq. (\ref{eq:Alpha})). However, these calculations depend on the calculation of $R[F32](n)$ and $r[F32](n)$. Then, when the IK-HSD module receives the input signals at every $n$-th instant the circuit shown in Figure \ref{fig:Fig_IMP_CI_Theta1} performs the calculation of $\theta_1^{HSD}[F32](n)$ in parallel with the generic circuits illustrated in Figures \ref{fig:Fig_IMP_CI_R} and \ref{fig:Fig_IMP_CI_rpequeno} which were implemented in FPGA to perform the calculation of $R[F32](n)$ and $r[F32](n)$ based on (\ref{eq:Rn}) and (\ref{eq:rn}).

\Figure[ht]()[width=0.85\linewidth]{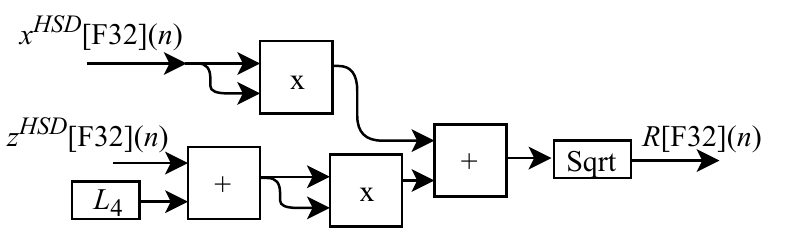}
{Proposed circuit to perform the calculation of $R[F32](n)$ (Eq. (\ref{eq:Rn})) - IK-HSD.\label{fig:Fig_IMP_CI_R}}

The circuit shown in Figure \ref{fig:Fig_IMP_CI_R} used to obtain $R[F32](n)$, is designed to work with two input signals $x^{HSD}[F32](n)$ and $z^{HSD}[F32](n)$ and one output signal. This design contains two multipliers, two adders, the $L_4$ constant and a sub-circuit called $Sqrt$, which was implemented in hardware to calculate the square root.

\Figure[ht]()[width=0.90\linewidth]{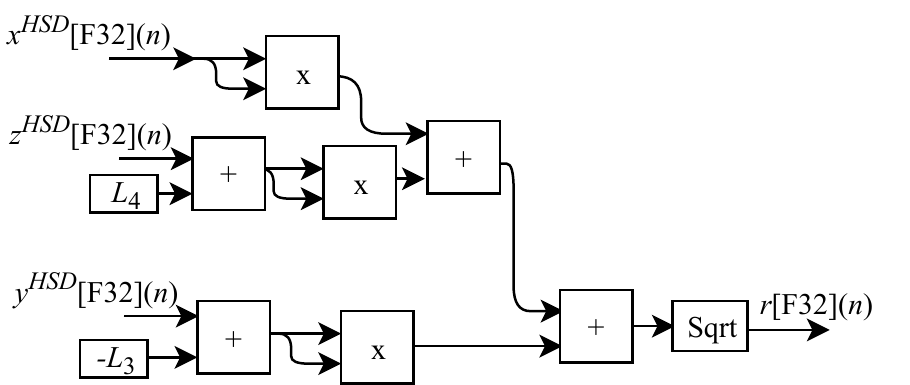}
{Proposed circuit to perform the calculation of $r[F32](n)$ (Eq. (\ref{eq:rn})) - IK-HSD.\label{fig:Fig_IMP_CI_rpequeno}}

The $r[F32](n)$ calculation is performed through the circuit shown in Figure \ref{fig:Fig_IMP_CI_rpequeno}. This circuit is designed to work with three input signals $x^{HSD}[F32](n)$, $y^{HSD}[F32](n)$ and $z^{HSD}[F32](n)$ and one output signal. The circuit consists of three multipliers, four adders, one inverter, the constants $L_3$ and $L_4$, and, again, the $Sqrt$ sub-circuit.

After the parallel processing of $\theta_1^{HSD}[F32](n)$, $R[F32](n)$ and $r[F32](n)$, the circuits responsible for calculating $\gamma[F32](n)$, $\beta[F32](n)$ and $\alpha[F32](n)$ are also executed in parallel through the FPGA implementations of the generic circuits illustrated in Figures \ref{fig:Fig_IMP_CI_Gamma}, \ref{fig:Fig_IMP_CI_Beta} and \ref{fig:Fig_IMP_CI_Alpha}. The value of $\gamma[F32](n)$ is obtained through the circuit shown in Figure \ref{fig:Fig_IMP_CI_Gamma} which is based on (\ref{eq:Gama}). The circuit is designed to work with an input signal $r[F32](n)$ and one output signal. It consists of five multipliers, two adder, one divisor, one TFB sub-circuit to calculate the arccosine and the constants $L_1$ and $L_2$.

\Figure[ht]()[width=0.95\linewidth]{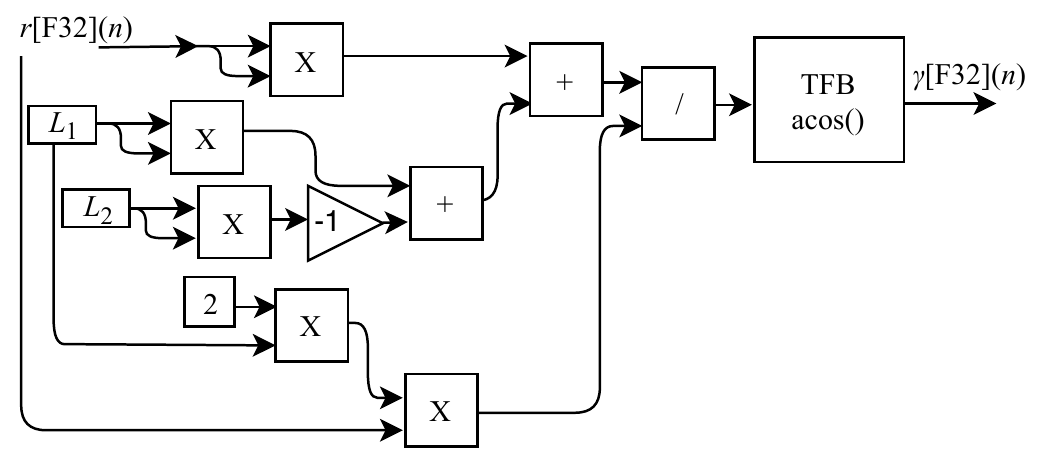}
{Proposed circuit to perform the calculation of $\gamma[F32](n)$ (Eq. (\ref{eq:Gama})) - IK-HSD.\label{fig:Fig_IMP_CI_Gamma}}

The circuit for obtaining $\beta[F32](n)$ illustrated in Figure \ref{fig:Fig_IMP_CI_Beta} is based on (\ref{eq:Beta}) and is designed to work with two input signals $y^{HSD}[F32](n)$ and $R[F32](n)$ and one output signal. The circuit is composed of one adder, one inverter, a TFB sub-circuit to perform the arctangent calculation and the $L_3$ constant.

\Figure[ht]()[width=0.7\linewidth]{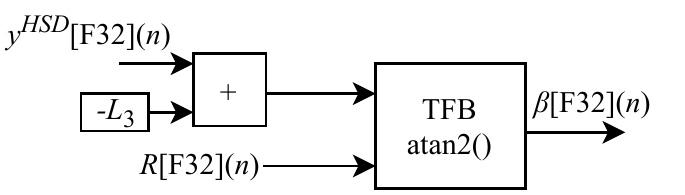}
{Proposed circuit to perform the calculation of $\beta[F32](n)$ (Eq. (\ref{eq:Beta})) - IK-HSD.\label{fig:Fig_IMP_CI_Beta}}

The value of $\alpha[F32](n)$ is obtained from the circuit shown in Figure \ref{fig:Fig_IMP_CI_Alpha} which is based on (\ref{eq:Alpha}) and is designed to work with an input signal $r[F32](n)$ and one output signal. The circuit is composed of five multipliers, two adders, one inverter, one divider, one TFB sub-circuit to perform the arccosine calculation and the constants $L_1$ and $L_2$.

\Figure[ht]()[width=0.95\linewidth]{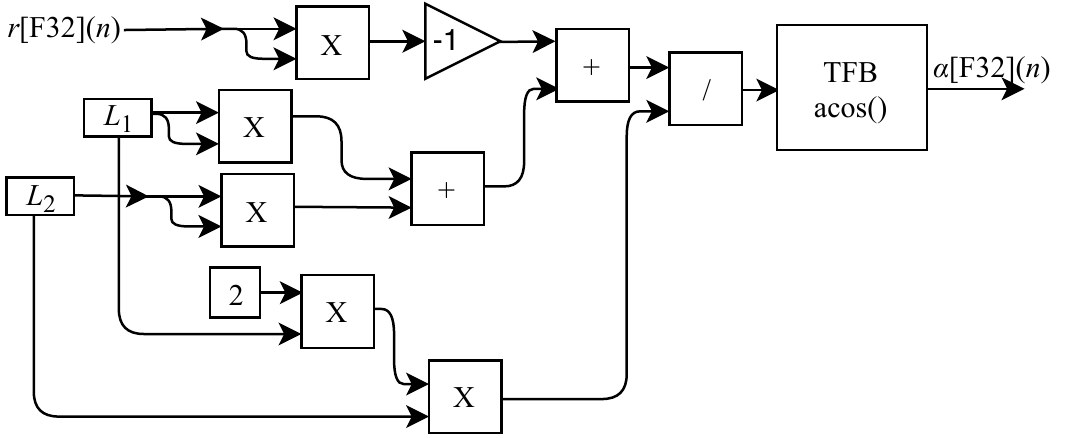}
{Proposed circuit to perform the calculation of $\alpha[F32](n)$ (Eq. (\ref{eq:Alpha})) - IK-HSD.\label{fig:Fig_IMP_CI_Alpha}}

To complete the process, after performing the calculations of $\beta[F32](n)$, $\gamma[F32](n)$ and $\alpha[F32](n)$, it is possible to obtain the $\theta_2^{HSD}[F32](n)$ and $\theta_3^{HSD}[F32](n)$ values in parallel through the circuits shown in Figures \ref{fig:Fig_IMP_CI_Theta2} and \ref{fig:Fig_IMP_CI_Theta3}.

\subsection{KINESTHETIC FEEDBACK FORCE (KFF-HMD)}

As illustrated in Figure \ref{fig:Fig_TactileModel_Detailed}, the hardware associated with the master device (HMD) implements the kinesthetic feedback force through the KFF-HMD module. Based on (\ref{eq:KinestheticFeedbackForce}), the KFF-HMD module was implemented in FPGA through the generic circuit illustrated in Figure \ref{fig:Fig_IMP_KFF_Module}. This circuit is composed of sub-circuits that correspond to parts of (\ref{eq:KinestheticFeedbackForce}). The sub-circuit called JM, described in (\ref{eq:Jacobian}), is responsible for calculating the Jacobian matrix. The KFF sub-circuit makes the relationship between the Jacobian matrix (JM) module and the force array from (\ref{eq:Force}).

\Figure[ht]()[width=0.85\linewidth]{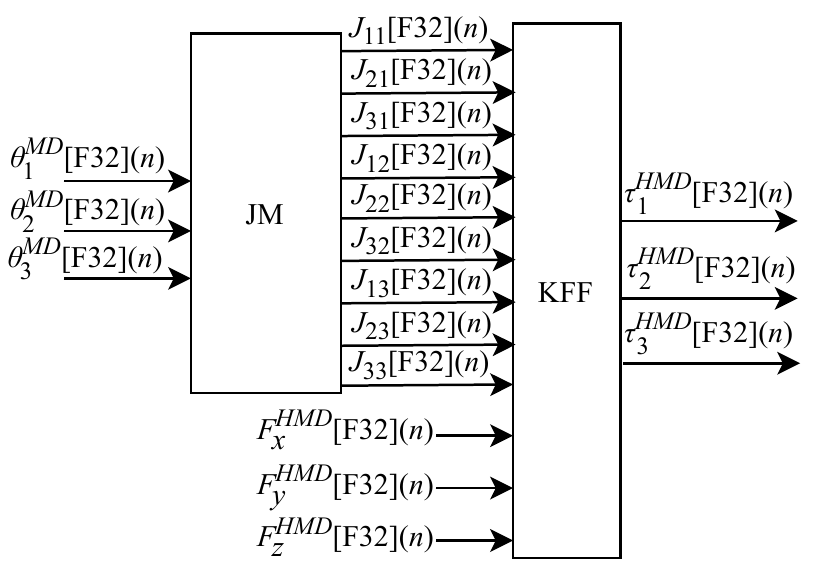}
{Proposed circuit to calculate kinesthetic feedback force (Eq. (\ref{eq:KinestheticFeedbackForce})) - KFF-HMD.\label{fig:Fig_IMP_KFF_Module}}

The circuit shown in Figure \ref{fig:Fig_IMP_KFF_Module} has the input signals $\theta_1^{MD}[F32](n)$ , $\theta_2^{MD}[F32](n)$ and $\theta_3^{MD}[F32](n)$ that are received from the master device (MD) and also the $F_x[F32](n)$, $F_y[F32](n)$ and $F_z[F32](n)$ signals that are received from the hardware associated to the slave device (HSD). The three output signals are: $\tau_1^{HMD}[F32](n)$, $\tau_2^{HMD}[F32](n)$ and $\tau_3^{HMD}[F32](n)$.

The JM module that represents the sub-circuit responsible for performing the Jacobian matrix calculation consists of nine elements: $J_{11}[F32](n)$, $J_{21}[F32](n)$, $J_{31}[F32](n)$, $J_{12}[F32](n)$, $J_{22}[F32](n)$, $J_{32}[F32](n)$, $J_{13}[F32](n)$, $J_{23}[F32](n)$ and $J_{33}[F32](n)$. The calculation of $J_{21}[F32](n)$ based on (\ref{eq:JM_21}) does not have an associated circuit since its value is $0$, i.e. $J_{21}[F32](n)=0$. Based on (\ref{eq:JM_11}), the algorithm for calculating $J_{11}[F32](n)$ was implemented in FPGA according to the generic circuit illustrated in Figure \ref{fig:Fig_IMP_JACOBIAN_J11}. The circuit was designed to work with three input signals and one output signal. It uses the constants $L_1$ and $L_2$ and has three TFB sub-circuits: two for performing the cosine calculation and one for obtaining the sine value.

\Figure[!h]()[width=0.9\linewidth]{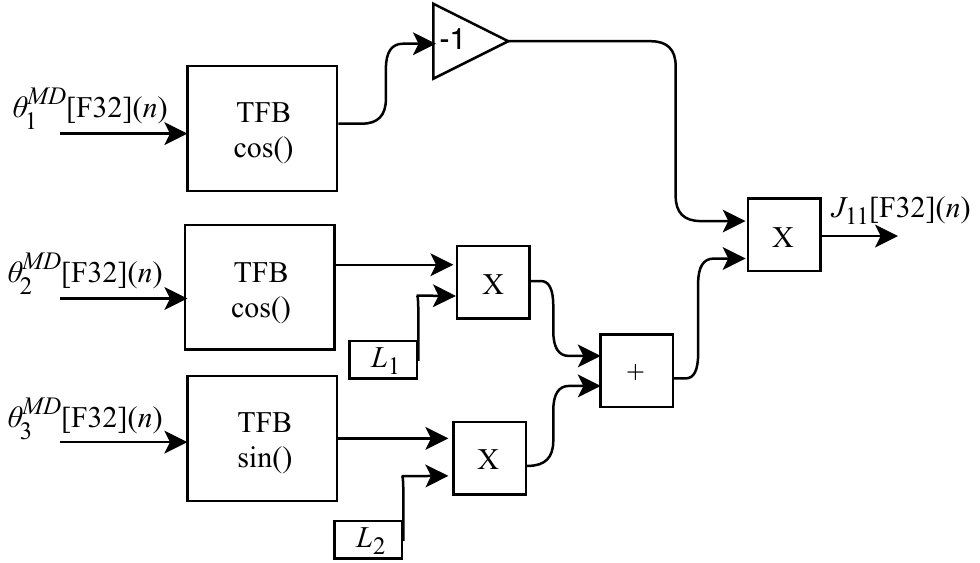}
{Proposed circuit to calculate the Jacobian matrix $J_{11}[F32](n)$ (Eq. (\ref{eq:JM_11})) - JM.\label{fig:Fig_IMP_JACOBIAN_J11}}

The calculation of $J_{31}[F32](n)$, based on (\ref{eq:JM_31}), was implemented in FPGA according to the generic circuit illustrated in Figure \ref{fig:Fig_IMP_JACOBIAN_J31}. The circuit was designed to work with three input signals and one output signal. The circuit has three TFB modules, two for sine calculation and one for cosine value and uses the $L_1$ and $L_2$ constants.

\Figure[!h]()[width=0.9\linewidth]{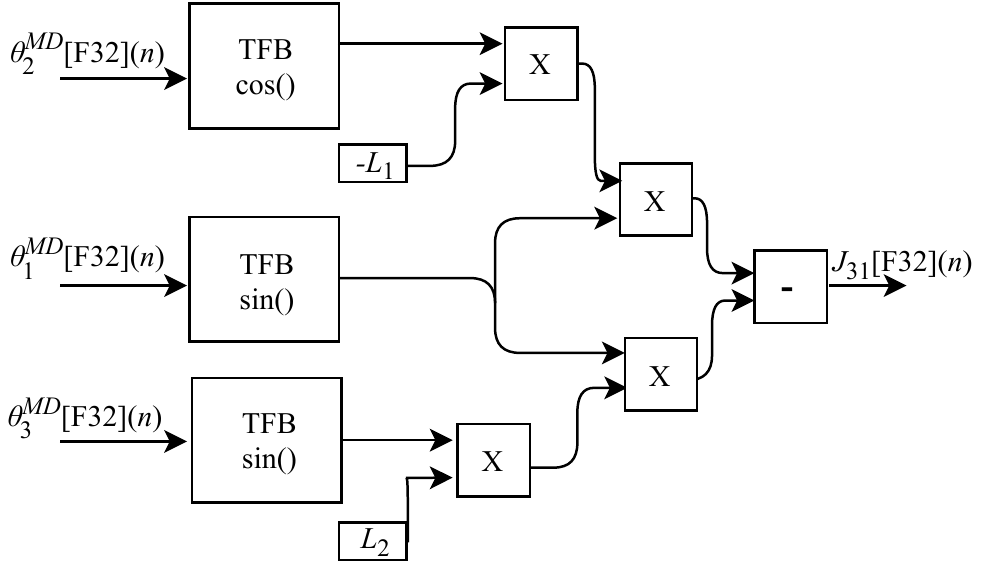}
{Proposed circuit to calculate the Jacobian matrix $J_{31}[F32](n)$ (Eq. (\ref{eq:JM_31})) - JM.\label{fig:Fig_IMP_JACOBIAN_J31}}

The generic circuit illustrated in Figure \ref{fig:Fig_IMP_JACOBIAN_J12} was implemented in FPGA to perform the calculation of $J_{12}[F32](n)$ and is based on (\ref{eq:JM_12}). The circuit was designed to work with two input signals and one output signal. The circuit has two TFB sub circuits to perform sine calculation and uses the $L_1$ constant.

\Figure[!h]()[width=0.85\linewidth]{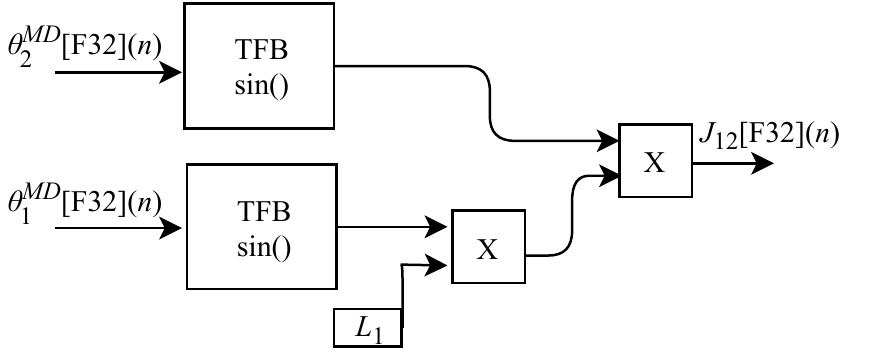}
{Proposed circuit to calculate the Jacobian matrix  $J_{12}[F32](n)$ (Eq. (\ref{eq:JM_12})) - JM.\label{fig:Fig_IMP_JACOBIAN_J12}}

Based on (\ref{eq:JM_22}), the algorithm for calculating $J_{22}[F32](n)$ was implemented in FPGA according to the generic circuit illustrated in Figure \ref{fig:Fig_IMP_JACOBIAN_J22}. The circuit was designed to work with one input signal and one output signal. The circuit has a TFB sub-circuit to perform cosine calculation and uses the constant $L_1$.

\Figure[!h]()[width=0.7\linewidth]{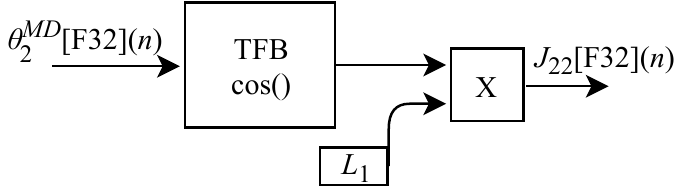}
{Proposed circuit to calculate the Jacobian matrix  $J_{22}[F32](n)$ (Eq. (\ref{eq:JM_22})) - JM.\label{fig:Fig_IMP_JACOBIAN_J22}}

The calculation of $J_{32}[F32](n)$ based on (\ref{eq:JM_32}) was implemented in  FPGA according to the generic circuit illustrated in Figure \ref{fig:Fig_IMP_JACOBIAN_J32}. The circuit was designed to work with two input signals and one output signal. In addition to the use of the constant $L_1$, the circuit has two TFB sub circuits, one for performing the cosine calculation and one for the sine.

\Figure[!h]()[width=0.85\linewidth]{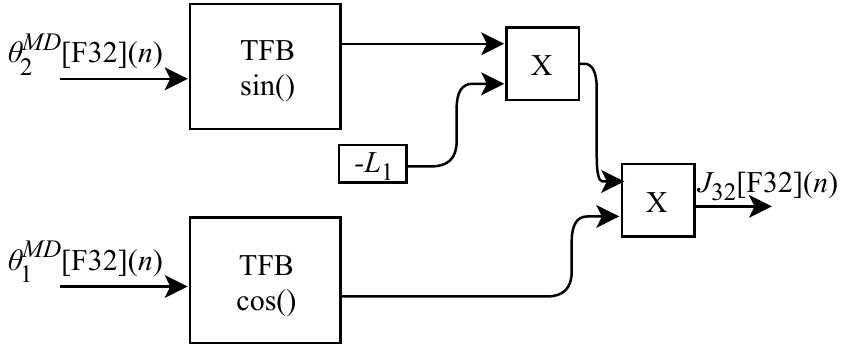}
{Proposed circuit to calculate the Jacobian matrix  $J_{32}[F32](n)$ (Eq. (\ref{eq:JM_32})) - JM.\label{fig:Fig_IMP_JACOBIAN_J32}}

The generic circuit illustrated in Figure \ref{fig:Fig_IMP_JACOBIAN_J13} was implemented in  FPGA to perform the calculation of $J_{13}[F32](n)$ and which is based on (\ref{eq:JM_13}). The circuit was designed to work with two inputs and one output signal. In addition to using the constant $L_2$, the circuit has two TFB sub circuits, one for performing cosine calculation and one for the sine.

\Figure[!h]()[width=0.85\linewidth]{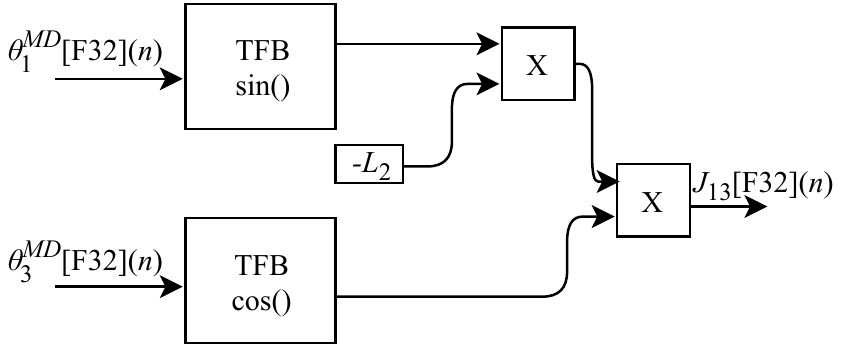}
{Proposed circuit to calculate the Jacobian matrix  $J_{13}[F32](n)$ (Eq. (\ref{eq:JM_13})) - JM.\label{fig:Fig_IMP_JACOBIAN_J13}}

Based on (\ref{eq:JM_23}), the algorithm for calculating $J_{23}[F32](n)$ was implemented in FPGA according to the generic circuit illustrated in Figure \ref{fig:Fig_IMP_JACOBIAN_J23}. The circuit was designed to work with one input signal and one output signal. The circuit contains a TFB sub-circuit to perform the sine calculation and uses the $L_2$ constant.

\Figure[!h]()[width=0.75\linewidth]{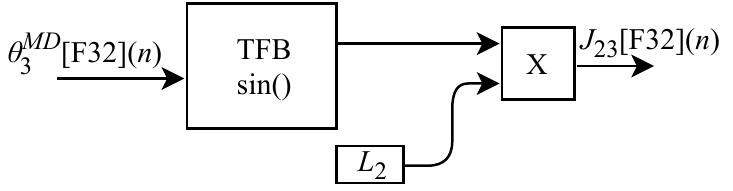}
{Proposed circuit to calculate the Jacobian matrix $J_{23}[F32](n)$ (Eq. (\ref{eq:JM_23})) - JM.\label{fig:Fig_IMP_JACOBIAN_J23}}

The calculation of $J_{33}[F32](n)$, based on (\ref{eq:JM_33}), was implemented in FPGA according to the generic  circuit illustrated in Figure \ref{fig:Fig_IMP_JACOBIAN_J33}. The circuit was designed to work with two input signals and one output signal. In addition to the use of constant $L_2$, the circuit has two TFB sub-circuits to perform the cosine calculation.

\Figure[!h]()[width=0.85\linewidth]{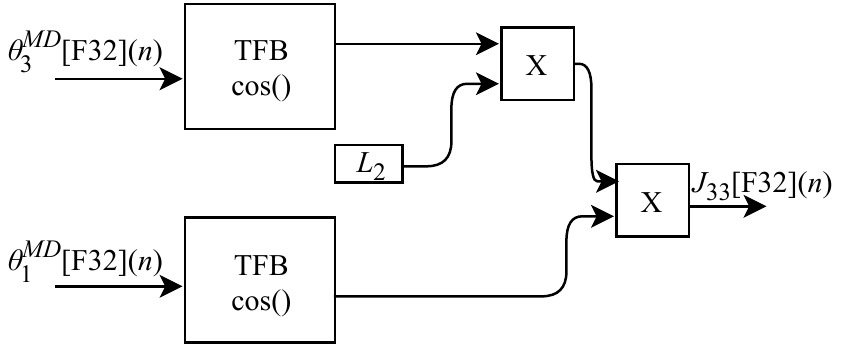}
{Proposed circuit to calculate the Jacobian matrix $J_{33}[F32](n)$ (Eq. (\ref{eq:JM_33})) - JM.\label{fig:Fig_IMP_JACOBIAN_J33}}

All displayed circuits related to the JM sub-circuits are calculated in parallel at each $n$-th instant. The results are then sent to the KFF module which also performs the calculations of $\tau_1^{HMD}[F32](n)$, $\tau_2^{HMD}[F32](n)$ and $\tau_3^{HMD}[F32](n)$ in parallel. The KF circuit shown in Figure \ref{fig:Fig_IMP_KFF_Module} is designed to work with twelve input signals and three output signals.

Based on (\ref{eq:Tau_1}), the algorithm for calculating $\tau_1^{HMD}[F32](n)$ was implemented in FPGA according to the generic circuit illustrated in Figure \ref{fig:Fig_IMP_KFF_Tau_1}. The circuit was designed to work with six inputs and one output.

\Figure[ht]()[width=0.8\linewidth]{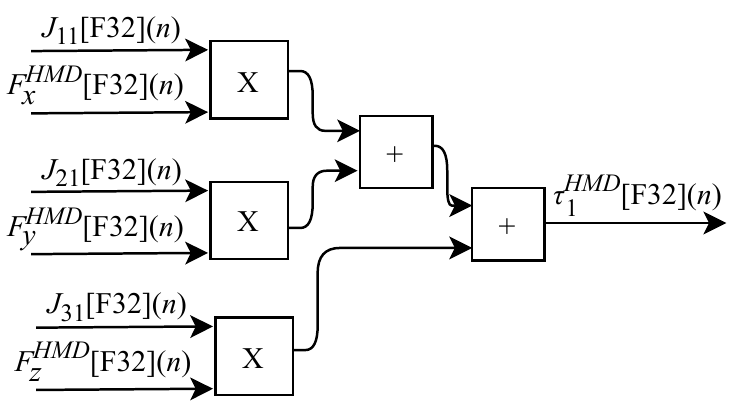}
{Proposed circuit to calculate the torque of the $\tau_1^{HMD}[F32](n)$ joint (Eq. (\ref{eq:Tau_1})) - KFF.\label{fig:Fig_IMP_KFF_Tau_1}}

The calculation of $\tau_2^{HMD}[F32](n)$ based on (\ref{eq:Tau_2}) was implemented in  FPGA according to the generic circuit illustrated in \ref{fig:Fig_IMP_KFF_Tau_2}. The circuit was designed to work with six inputs and one output.

\Figure[ht]()[width=0.8\linewidth]{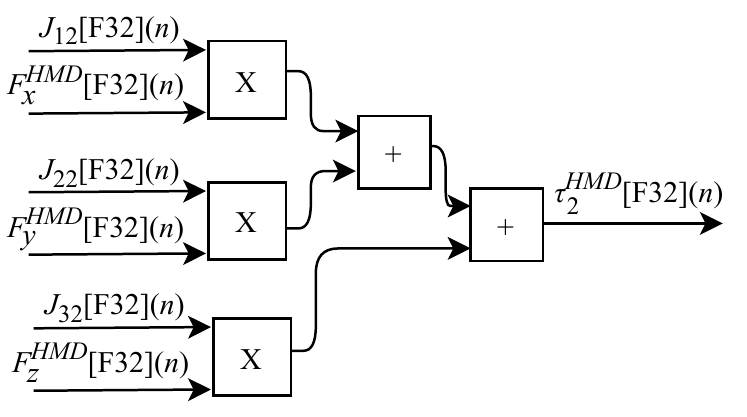}
{Proposed circuit to calculate the torque of the $\tau_2^{HMD}[F32](n)$ joint (Eq. (\ref{eq:Tau_2})) - KFF.\label{fig:Fig_IMP_KFF_Tau_2}}

The generic circuit illustrated in Figure \ref{fig:Fig_IMP_KFF_Tau_3} has been implemented in  FPGA to perform the calculation of $\tau_3^{HMD}[F32](n)$ and it is based on (\ref{eq:Tau_3}). The circuit was designed to work with six inputs and one output.

\Figure[ht]()[width=0.8\linewidth]{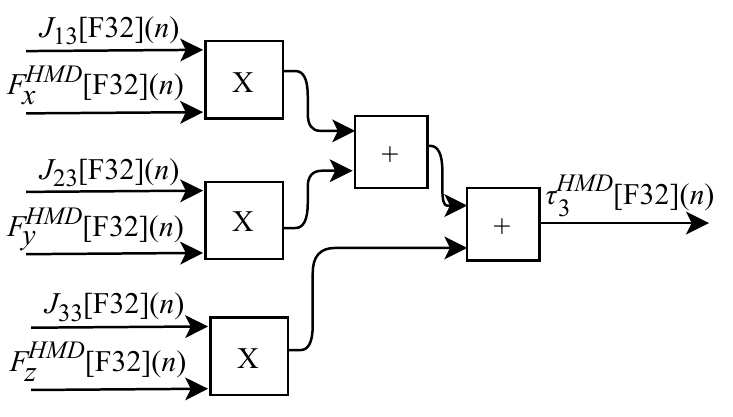}
{Proposed circuit to calculate the torque of the $\tau_3^{HMD}[F32](n)$ joint (Eq. (\ref{eq:Tau_3})) - KFF.\label{fig:Fig_IMP_KFF_Tau_3}}

\subsection{FEEDBACK FORCE (FBF-HSD)}

As illustrated in Figure \ref{fig:Fig_TactileModel_Detailed} the hardware associated with the slave device (HSD) implements the feedback force via the FBF-HSD module. The FPGA-implemented circuit of the FBF-HSD module is designed to work with six input signals and three output signals. Among the six input variables, $x^{OBJ}[F32](n)$, $y^{OBJ}[F32](n)$ and $z^{OBJ}[F32](n)$ represent the spatial position of the closest object to the SD tool and the other three $x^{ENV}[F32](n)$, $y^{ENV}[F32](n)$ and $z^{ENV}[F32](n)$ represent the spatial position of the SD tool in the ENV module. The three outputs $F_x^{HSD}[F32](n)$, $F_y^{HSD}[F32](n)$ and $F_z^{HSD}[F32](n)$ represent the touch of the tool on the object. The variables $h_x$, $h_y$ and $h_z$ represent the elasticity coefficients associated with the object. All FBF-HSD module calculations are performed in parallel.

Based on (\ref{eq:FF_Fx}), the algorithm used for calculating $F_x^{HSD}[F32](n)$ was implemented in FPGA according to the generic circuit illustrated in Figure \ref{fig:Fig_IMP_FB_Fx}. The circuit was designed to work with two inputs signals $x^{OBJ}[F32](n)$ and $x^{ENV}[F32](n)$ and one variable $h_x$.

\Figure[ht]()[width=0.65\linewidth]{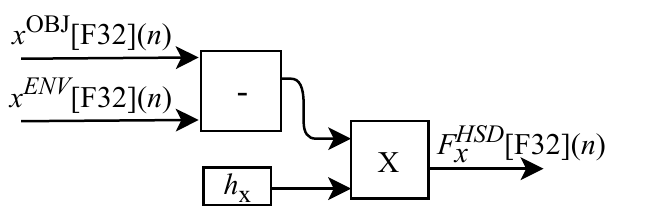}
{Proposed circuit to calculate the feedback force $F_x^{HSD}[F32](n)$ (Eq. (\ref{eq:FF_Fx})) - FBF-HSD.\label{fig:Fig_IMP_FB_Fx}}

The calculation of $F_y^{HSD}[F32](n)$, based on (\ref{eq:FF_Fy}), was implemented in FPGA according to the generic circuit illustrated in Figure \ref{fig:Fig_IMP_FB_Fy}. The circuit was designed to work with two input signals $y^{OBJ}[F32](n)$ and $y^{ENV}[F32](n)$ and one variable $h_y$.
 
\Figure[ht]()[width=0.65\linewidth]{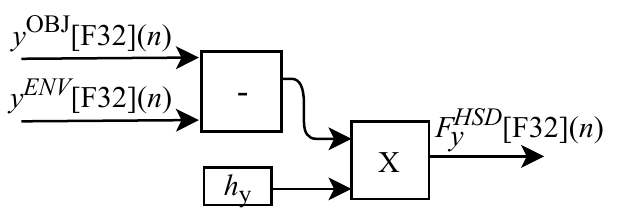}
{Proposed circuit to calculate the feedback force $F_y^{HSD}[F32](n)$ (Eq. (\ref{eq:FF_Fy})) - FBF-HSD.\label{fig:Fig_IMP_FB_Fy}}

The generic circuit shown in Figure \ref{fig:Fig_IMP_FB_Fz} was implemented in FPGA to perform the calculation of $F_z^{HSD}[F32](n)$ and it is based on (\ref{eq:FF_Fz}). The circuit was designed to work with two input signals $z^{OBJ}[F32](n)$ and $z^{ENV}[F32](n)$ and one variable $h_z$.

\Figure[ht]()[width=0.65\linewidth]{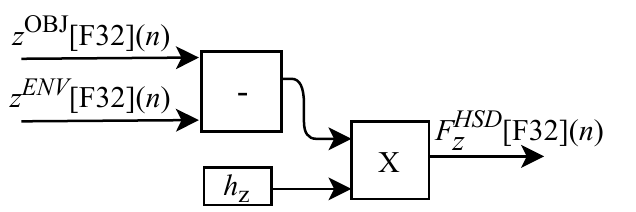}
{Proposed circuit to calculate the feedback force $F_z^{HSD}[F32](n)$ (Eq. (\ref{eq:FF_Fz})) - FBF-HSD.\label{fig:Fig_IMP_FB_Fz}}

\section{Results}

The entire tactile internet model infrastructure presented in Figure \ref{fig:Fig_TactileModel_Detailed} was implemented with the purpose of validating the FPGA hardware implementation. A spatial trajectory that represents the data sent by the OP through the $\mathbf{a}(n)$ (Eq. (\ref{eq:Signal_A})) signal was created to validate the entire developed environment.

The created trajectory performs a variation in all of the three angles of the MD articulation. (Figure \ref{fig:OmniPhantom}). For this, it was first considered that the MD is in the initial angular position expressed as $\theta^{MD}_1(0)=0$, $\theta^{MD}_2(0)=0$ and $\theta^{MD}_3(0)=0$, which corresponds to the spatial position $x^{OP}(0)=0$, $y^{OP}(0)=-0.107$ and $z^{OP}(0)=-0.035$ of the tool as illustrated in Figure \ref{fig:Fig_Trajectory}. Initially, the first joint is moved to $\theta^{MD}_1(vn)=pi/2$ where $v$ represents a quantity of samples that is equal to 4 seconds, thus resulting in the position $x^{OP}(vn)=-0.132$, $y^{OP}(vn)=-0.107$ and $z^{OP}(vn)=-0.167$. Then, the second joint is moved to $\theta^{MD}_2(vn)=pi/4$ which results in the position $x^{OP}(vn)=-0.093$, $y^{OP}(vn)=-0.013$ and $z^{OP}(vn)=-0.167$ and, finally, the third joint moves up to $\theta^{MD}_3(vn)=pi/4$, thus resulting in the $x^{OP}(vn)=-0.186$, $y^{OP}(vn)=0.025$ and $z^{OP}(vn)=-0.167$ position. The path created is within the limits of the device workspace and takes a total time of $t_1=12$ seconds of which $4$ seconds are used to perform the movement of each joint.

\Figure[ht]()[width=0.9\linewidth]{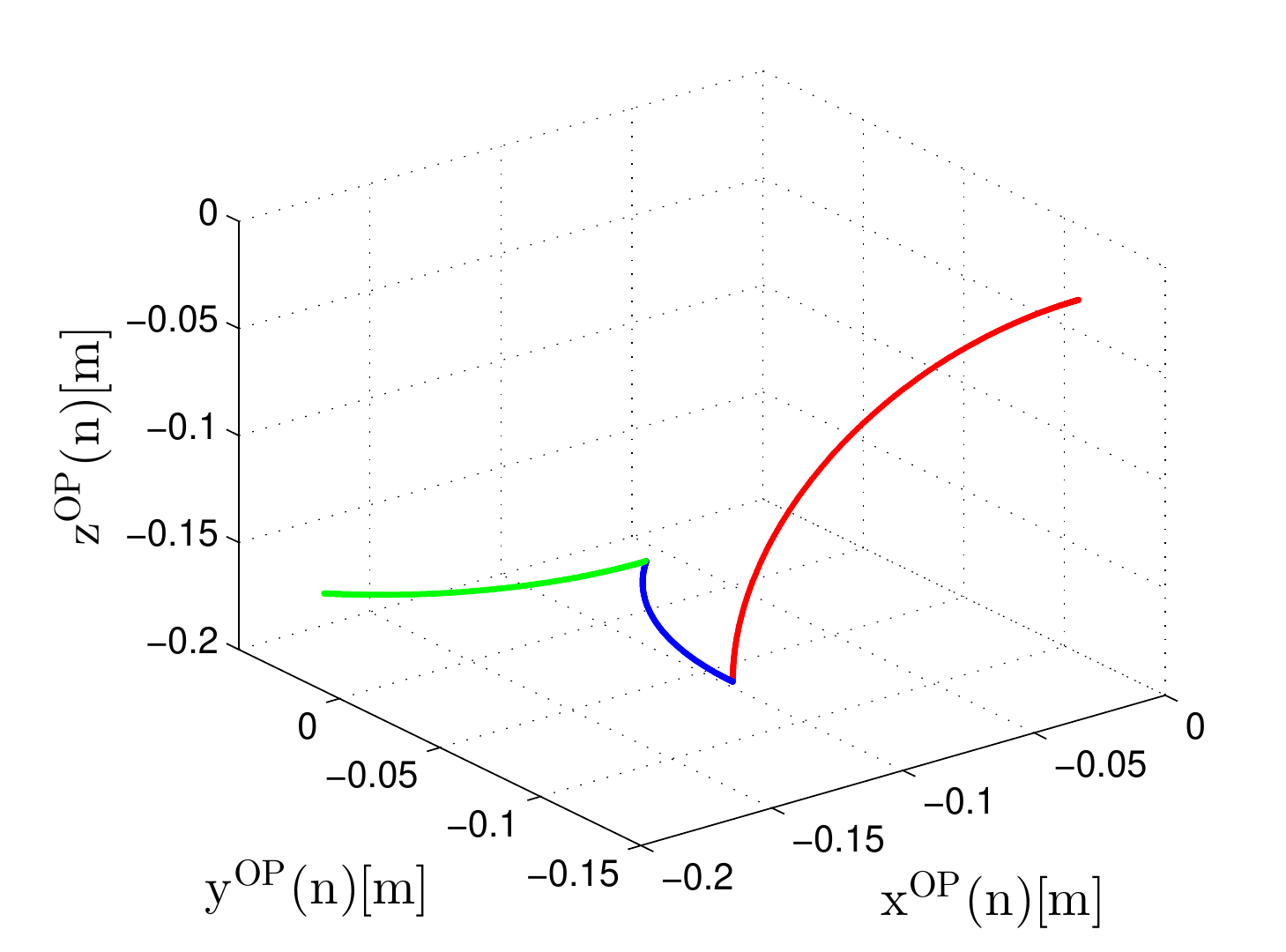}
{Trajectory used to validate hardware modules.\label{fig:Fig_Trajectory}}

In an effort to validate the circuits from the implemented modules in FPGA, equivalent software models were used to compare the results of both implementations. The software models use a $32$-bit floating point format while the hardware modules run a parallel implementation with a hybrid representation which uses both a $32$-bit floating point and a fixed point representation in different parts of the proposed architecture, as presented in Section \ref{sec:Implementation}. In all scenarios, the signal sampling rate (or throughput) was $R_s=\frac{1}{t_s}$ (samples per second), where $t_s$ is the time between the $n$-th samples.

From the experimental results, the mean square error (MSE) between the software model and the hardware implementation proposed by this work was calculated using the MSE which can be expressed as
\begin{equation}\label{eq:erroMedio}
MSE = \frac{1}{Q} \sum_{n=0}^{Q-1} (M^{SW}[F32](n) - M[F32](n))^2,
\end{equation}
where $Q$ represents the number of tested samples, $M^{SW}[F32](n)$ corresponds to the variables of the software model and $M[F32](n)$ corresponds to the variables of the model implemented in FPGA.

The quantity of tested samples for the results presented here is $Q=1200$, which correspond to the quantity of samples of the generated trajectory. The variables that correspond to the hardware model $M[F32](n)$ vary according to the module in which it was implemented. In the case of forward kinematics, as the FK-HMD and FK-HSD modules have the same implementation, the values corresponding to the variables $x[F32](n)$, $y[F32](n)$ and $z[F32](n)$ change according to the respective module. For the FK-HMD module, these variables correspond to $x^{HMD}[F32](n)$, $y^{HMD}[F32](n)$ and $z^{HMD}[F32](n)$ and for the FK-HSD module the same variables correspond to $x^{ENV}[F32](n)$, $y^{ENV}[F32](n)$ and $z^{ENV}[F32](n)$ as presented in Section \ref{sec:Implementation}. For inverse kinematics, the variables $M[F32](n)$ of the IK-HSD module correspond to $\theta_1^{HSD}[F32](n)$, $\theta_2^{HSD}[F32](n)$ and $\theta_3^{HSD}[F32](n)$. For the kinesthetic feedback force, the variables $M[F32](n)$ of the KFF-HMD module correspond to $\tau^{HMD}_1[F32](n)$, $\tau^{HMD}_2[F32](n)$ and $\tau^{HMD}_3[F32](n)$. For the feedback force, the variables $M[F32](n)$ of the FBF-HSD module correspond to $F_x^{HSD}[F32](n)$, $F_y^{HSD}[F32](n)$ and $F_z^{HSD}[F32](n)$. And finally, in the MSE equation the $M^{SW}[F32](n)$ corresponds to the same variables as the software-implemented model.

Table \ref{tab:MSE_FloatingPoint} shows the mean square error between the software models and the hardware ones proposed in this paper. The obtained MSE-related results prove to be noteworthy, showing that the forward kinematics (FK-HMD and FK-HSD), inverse kinematics (IK-HSD), kinesthetic feedback force (KFF-HMD) and feedback force (FBF-HSD) modules had an acceptable response, even when using a hybrid representation, compared to the software model that uses a floating point representation. It can be observed that for the variables of the FK-HMD and FK-HSD modules the error was in the range of $10^{-08}$, for the IK-HSD module the error was of $10^{-06}$, for the variables of the KFF-HMD module the error was of $10^{-07}$ and for the FBF-HSD module the error was in the range of $10^{-16}$. These values demonstrate that the FPGA implementations presented an equivalent behavior to the software models.
 
\begin{table}[ht]
	\caption{Mean squared error (MSE) results for floating-point implementation.}
	\label{tab:MSE_FloatingPoint}
	\small
	\centering
	\begin{tabular}{ccc}
		\toprule[\heavyrulewidth]
		\textbf{Module} & \textbf{Variable} & \textbf{MSE} \\

		\cmidrule(r){1-3}			
		\multirow{3}{*}{\begin{tabular}{c} FK-HMD \end{tabular}} 
		& $x^{HMD}[F32](n)$ & \multirow{1}{*}{$2.333 \times 10^{-8}$} \\
		& $y^{HMD}[F32](n)$ & \multirow{1}{*}{$8.316 \times 10^{-9}$}  \\
		& $z^{HMD}[F32](n)$ & \multirow{1}{*}{$1.656 \times 10^{-8}$} \\	
		
		\cmidrule(r){1-3}
		\multirow{3}{*}{\begin{tabular}{c} KFF-HMD \end{tabular}} 
		& $\tau^{HMD}_1[F32](n)$ & \multirow{1}{*}{$1.467 \times 10^{-7}$} \\
		& $\tau^{HMD}_2[F32](n)$ & \multirow{1}{*}{$5.207 \times 10^{-9}$}  \\
		& $\tau^{HMD}_3[F32](n)$ & \multirow{1}{*}{$3.350 \times 10^{-7}$} \\	
		
		\cmidrule(r){1-3}			
		\multirow{3}{*}{\begin{tabular}{c} FK-HSD \end{tabular}} 
		& $x^{ENV}[F32](n)$ & \multirow{1}{*}{$2.333 \times 10^{-8}$} \\
		& $y^{ENV}[F32](n)$ & \multirow{1}{*}{$8.316 \times 10^{-9}$}  \\
		& $z^{ENV}[F32](n)$ & \multirow{1}{*}{$1.656 \times 10^{-8}$} \\	
			
		\cmidrule(r){1-3}
		\multirow{3}{*}{\begin{tabular}{c} IK-HSD \end{tabular}} 
		&  $\theta_1^{HSD}[F32](n)$ & \multirow{1}{*}{$3.731 \times 10^{-6}$} \\
		&  $\theta_2^{HSD}[F32](n)$ & \multirow{1}{*}{$2.847 \times 10^{-6}$}  \\
		&  $\theta_3^{HSD}[F32](n)$ & \multirow{1}{*}{$2.702 \times 10^{-6}$} \\	
			
		\cmidrule(r){1-3}
		\multirow{3}{*}{\begin{tabular}{c} FBF-HSD \end{tabular}} 
		& $F_x^{HSD}[F32](n)$ & \multirow{1}{*}{$2.437 \times 10^{-16}$} \\
		& $F_y^{HSD}[F32](n)$ & \multirow{1}{*}{$1.731 \times 10^{-16}$}  \\
		& $F_z^{HSD}[F32](n)$ & \multirow{1}{*}{$3.360 \times 10^{-16}$} \\	
			
		\hline        
	\end{tabular}
\end{table}

In a hardware implementation, it is important to analyze some requirements post-synthesis such as available hardware usage and the execution time. In the case of FPGAs, the resources are measured through the use of lookup tables (LUTs), Registers and Digital Signal Processors (DSPs) units, to name a few. After validating the hardware-implemented models, synthesis results were obtained using the implementation designed for an FPGA Xilinx Virtex 6 XC6VLX240T-1FF1156. The used Virtex 6 FPGA has $37\text{,}680$ slices that group $301\text{,}440$ flip-flops, $150\text{,}720$ logical cells that can be used to implement logical functions or memories, and $768$ DSP cells with multipliers and accumulators.

Table \ref{tab:ResultsSynthesis_FloatingPoint} presents the post-synthesis results related to hardware occupancy, sampling rate, and throughput for the modules FK-HMD, KFF-HMD, FK-HSD, IK-HSD, and FBF-HSD. The first column shows the name of the module, the next three columns called registers, LUTs and multipliers represent the amounts of resources used in the FPGA. The column register represents the number of flip-flops that were used, followed by the total percentage used. The column LUTs represents the number of LUTs that were used, followed by the total percentage used. And the column multipliers represents the number of DPS48 internal multipliers that were used, followed by the total percentage used. The $t_s$ column represents the sampling rate in nanoseconds that was obtained for each hardware module. Finally, the $R_s$ column displays throughput ($R_s=\frac{1}{t_s}$) values in mega-samples per second for the hardware modules.

\begin{table*}[ht]
	\caption{Hardware occupancy, sampling rate and throughput results for floating-point format.}
	\label{tab:ResultsSynthesis_FloatingPoint}
	\small
	\centering
	\begin{tabular}{ccccccc}
		\toprule[\heavyrulewidth]
		\begin{tabular}{c} \textbf{Module} \\ \textbf{Name}\end{tabular} & 
		\begin{tabular}{c} \textbf{Registers} \\ \textbf{(Flip-Flops)} \end{tabular} &
		\begin{tabular}{c} \textbf{LUTs} \end{tabular} &
		\begin{tabular}{c} \textbf{Multipliers} \\ \textbf{(DSP48)} \end{tabular} &
		\begin{tabular}{c} $t_s$  \\ (ns) \end{tabular} &
		\begin{tabular}{c} $R_s$  \\ (MSps) \end{tabular} \\
		\midrule
		\begin{tabular}{c} FK-HMD \end{tabular} & $3\text{,}041$ ($1\text{.}01$\%) & $8\text{,}008$ ($5\text{.}31$\%) & $11$ ($1\text{.}43$\%) & $47$ &  $21\text{.}27$\\ 
		\rowcolor[gray]{.9}	\begin{tabular}{c} KFF-HMD \end{tabular} & $3\text{,}113$ ($1\text{.}03$\%) & $12\text{,}251$ ($8\text{.}13$\%) & $48$ ($6\text{.}25$\%) & $70$ &  $14\text{.}28$\\ 
		\begin{tabular}{c} FK-HSD \end{tabular} & $3\text{,}041$ ($1\text{.}01$\%) & $8\text{,}008$ ($5\text{.}31$\%) & $11$ ($1\text{.}43$\%) & $47$ &  $21\text{.}27$\\  
		\rowcolor[gray]{.9}	\begin{tabular}{c} IK-HSD \end{tabular} & $3\text{,}149$ ($1\text{.}04$\%) & $14\text{,}107$ ($9\text{.}36$\%) & $27$ ($3\text{.}52$\%)& $218$ &  $4\text{.}58$\\  
		\begin{tabular}{c} FBF-HSD \end{tabular} & $323$ ($0\text{.}11$\%) & $1\text{,}236$ ($0\text{.}82$\%) & $9$ ($1\text{.}17$\%) & $21$ &  $47\text{.}61$\\  
		
		\bottomrule[\heavyrulewidth] 
	\end{tabular}
\end{table*}

The synthesis results presented in Table \ref{tab:ResultsSynthesis_FloatingPoint} show that the resources used for the FK-HMD and FK-HSD modules were the same. This means that each module, individually, used a percentage of $1\text{.}01$\% which is equivalent to $3\text{,}041$ of the available hardware resources for the registers, was used $5\text{.}31$\% with LUTs, and $1\text{.}43$\% for embedded multipliers DSP48. The IK-HSD module had a hardware percentage consumption of $1\text{.}04$\% for registers, $9\text{.}36$\% for LUTs and $3\text{.}52$\% for multipliers. The KFF-HMD module had a consumption of $1\text{.}03$\%, $8\text{.}13$\% and $6\text{.}25$\% for registers, LUTs and multipliers, respectively. Finally, the FBF-HSD module used a percentage of $0\text{.}11$\% for registers, $0\text{.}82$\% for LUTs and $1\text{.}17$\% for multipliers. 

Based on data presented in Table \ref{tab:ResultsSynthesis_FloatingPoint}, the HMD modules (FK-HMD and KFF-HMD) that is associated with the MD device has consumed $6\text{,}154$ ($2\text{.}04$\%) for register, $20\text{,}259$ ($13\text{.}44$\%) for LUTs and $59$ ($7\text{.}68$\%) for multipliers. In the case of hardware associated with the SD device, the HSD modules (FK-HSD, IK-HSD and FBF-HSD) had consumed $6\text{,}513$ ($2\text{.}16$\%) for register, $23\text{,}351$ ($15\text{.}49$\%) for LUTs and $47$ ($6\text{.}12$\%) for multipliers.

The hardware resources consumed by the HMD hardware modules and the HSD hardware modules were very low. Even if all modules are implemented in single hardware, the consumption remains low. The total sum of hardware resources used in the FPGA by all modules (FK-HMD, KFF-HMD, FK-HSD, IK-HSD and FBF-HSD) was: $12\text{,}667$ ($4\text{.}20$\%) for register, $43\text{,}610$ ($28\text{.}93$\%) for LUTs and $106$ ($13\text{.}80$\%) for multipliers. The low hardware resources consumption demonstrates that the proposed implementations take up little hardware space in the FPGA which allows other separate implementations to be used concomitantly.

As per Table \ref{tab:ResultsSynthesis_FloatingPoint}, the throughput values, $R_s$, obtained were significant. Values of $21.27 \, \text{MSps}$ for the FK-HMD and FK-HSD modules, $4.58 \, \text{MSps}$ for the IK-HSD module, $14.28 \, \text{MSps}$ for the KFF-HMD module and $47.61 \, \text{MSps}$ for the FBF-HSD module were achieved. These results enable critical applications that demand strict time constraints, as is the case with tactile internet applications.

In Table \ref{tab:Speedup_FloatingPoint}, it is possible to see the speedup obtained in relation to latency time constraints. The first column presents the latency constraints of $1 \, \text{ms}$ and $10 \, \text{ms}$ that are presented in the literature. The second column shows the minimum latency values that are required for the application to function normally. The third column shows the latency related with the hardware implementation presented here.

\begin{table}[ht]
	\caption{Hardware speedup related to the time limits for the $1 \, \text{ms}$ and $10 \, \text{ms}$ latency constraints.} 
	\label{tab:Speedup_FloatingPoint}
	\small
	\centering
	\begin{tabular}{cccc}
		\toprule[\heavyrulewidth]
		\begin{tabular}{c} \textbf{Time }  \\ \textbf{Restriction} \end{tabular} &
		\begin{tabular}{c} \textbf{Latency}  \\ \textbf{Limit} \end{tabular} &
		\begin{tabular}{c} $t_{\text{hardware}}$ \end{tabular} &
		\begin{tabular}{c} \textbf{Speedup} \end{tabular} \\
		\midrule
		$1 \, \text{ms}$ & $37.5 \, \mu\text{s}$ & $403 \, \text{ns}$  & $93\,\times$ \\
		$10 \, \text{ms}$ & $375 \, \mu\text{s}$ & $403 \, \text{ns}$ & $930\,\times$ \\
		\bottomrule[\heavyrulewidth] 
	\end{tabular}
\end{table}

The $1 \, \text{ms}$ restriction corresponds to the maximum latency limit of $37.5 \, \mu\text{s}$ for acceptable hardware performance. For the $10 \, \text{ms}$ constraint, the maximum limit is $375 \, \mu\text{s}$. The value $t_{\text{hardware}}$ that is presented in Table \ref{tab:Speedup_FloatingPoint} and according to (\ref{timeHardware}), corresponds to the sum of the latencies of the five implemented modules (Table \ref{tab:ResultsSynthesis_FloatingPoint}), two modules are associated with the MD device (FK-HMD and KFF-HMD) and three modules are associated with the SD device (FK-HSD, IK-HSD, and FBF-HSD).

Thus, the presented value of $403 \, \text{ns}$ in Table \ref{tab:Speedup_FloatingPoint} corresponds to the sum of the two modules related to the master component, which has a total of $117 \, \text{ns}$ of which $47 \, \text{ns}$ come from the FK-HMD module and $70 \, \text{ns}$ from the KFF-HMD module together with the sum of the three modules referring to the slave component, which has a total of $286 \, \text{ns}$ of which $47 \, \text{ns}$ derives from the FK-HSD module, $218 \, \text{ns}$ from IK-HSD and $21 \, \text{ns}$ from the FBS-HSD module. So for the $1 \, \text{ms}$ constraint, the implementation presented a $93\,\times$ speedup relative to the $37.5 \, \mu\text{s}$, and for the $10 \, \text{ms}$ constraint, the speedup was  $930\,\times$ relatives to the $375 \, \mu\text{s}$ limit.

The sample rates resulted from the five modules that were implemented in this work were notably fast. The values obtained contributed to the hardware meeting the time constraint limits required in a tactile internet environment. Hardware latency showed values significantly below the required constraints, as shown in Table \ref{tab:Speedup_FloatingPoint}. These values are well below the $30$\% presented in the literature and due to the fact that the communication medium demands $70$\% of application latency, this value can be increased as the latency of hardware devices showed to be significantly low. In other words, it can be said that the remaining latency not spent on the hardware devices can be consumed in the network.

It is important to remember that in a more complex tactile internet environment, there are several others more algorithms to be implemented in hardware such as prediction algorithms, dynamic control, AI based techniques, etc. However, as the proposed implementations present low hardware resource consumption, other necessary modules, as the ones previously mentioned, could also be implemented in the same shared hardware since resources would still be available.

Table \ref{tab:ResultsComparisons_FloatingPoint} presents comparisons of the results obtained by the proposed implementation of this work with equivalent results found in works from the state of the art. The first column indicates the references of related works. The next two columns show the used FPGA platform and the amount of degrees of freedom of the used device. The fourth column presents the type of numerical representation used in the implementation and, finally, the last four columns present the times obtained by each reference for latency added by the forward kinematics (FK), inverse kinematics (IK), the kinesthetic force feedback (KFF) and feedback force (FBF) modules, respectively.

\begin{table*}[!h]
	\caption{Comparative table with state of the art works.} 
	\label{tab:ResultsComparisons_FloatingPoint}
	\small
	\centering
	\begin{tabular}{cccccccc}
		\toprule[\heavyrulewidth]
		
		\begin{tabular}{c} \textbf{Reference} \end{tabular} & 
		\begin{tabular}{c} \textbf{Device} \end{tabular} &
		\begin{tabular}{c} \textbf{DoF} \end{tabular} &
		\begin{tabular}{c} \textbf{Data type} \end{tabular} &
		\begin{tabular}{c} \textbf{FK} \end{tabular} &
		\begin{tabular}{c} \textbf{IK} \end{tabular} &
		\begin{tabular}{c} \textbf{KFF} \end{tabular} &
		\begin{tabular}{c} \textbf{FBF} \end{tabular} \\ 
		\midrule
		This work & Virtex 6 & 3 & Floating P. & $47  \, \text{ns}$ & $ 218 \, \text{ns}$ & $ 70 \, \text{ns}$ & $ 21 \, \text{ns}$ \\
		\rowcolor[gray]{.9}\cite{2010_fpgaKinematicRoboticManupulator} & Virtex 2 & 5 & Floating P. & $ 1240 \, \text{ns}$ & - & - & -\\
	    \cite{2013_fpgaParallelRobotKinematic} & Cyclone IV & 3 & Floating P. & - & $ 143000 \, \text{ns}$ & - & -\\
		\rowcolor[gray]{.9} \cite{fpga_2014_fixedPointKinematics} & Unknown & 6 & Fixed P. & $ 3000 \, \text{ns} $  & $ 4500 \, \text{ns} $ & - & -\\
		\cite{2013_fpgaKinematicsCordic} & Cyclone IV & 10 & Fixed P. & - & $ 440 \, \text{ns}$ & - & -\\
		\rowcolor[gray]{.9} \cite{2015_fpgaKinematicArticuledRobot} & Cyclone IV & 5 & Fixed P. & $ 680 \, \text{ns} $ & $940 \, \text{ns} $ & - & -\\
		\cite{2017_fpga_SurgicalRobotKinematic} & Artix 7 & 3 & Fixed P. & \multicolumn{2}{c}{ $ 2000 \, \text{ns} $ } & - & -\\
		\bottomrule[\heavyrulewidth] 
	\end{tabular}
\end{table*}

As described in Table \ref{tab:ResultsComparisons_FloatingPoint}, a hardware model for calculating the forward kinematics of a $5$-DoF device is presented in \cite{2010_fpgaKinematicRoboticManupulator}. The proposed hardware was implemented using a $32$-bit floating-point representation. The total time to perform the calculations was $1240 \, \text{ns}$. Comparing to the forward kinematics (FK) implementation using $32$-bit floating-point proposed by this work, the speedup was $26.38\,\times$ over the model presented in \cite{2010_fpgaKinematicRoboticManupulator}.

The work presented in \cite{2013_fpgaParallelRobotKinematic} shows the results of an implementation of the inverse kinematics module using floating-point 32-bit representation. The kinematic model was designed to work with a $3$-DoF device, and the time required to calculate is $143000 \, \text{ns}$.  When compared with the proposal of inverse kinematics (IK) presented in this work, which uses $32$-bit floating-point representation, this implementation presented a speedup of $655.96\,\times$ over in relation to the model proposed by \cite{2013_fpgaParallelRobotKinematic}.

The kinematics models presented in \cite{fpga_2014_fixedPointKinematics} described in Table \ref{tab:ResultsComparisons_FloatingPoint}, presented data regarding the forward and inverse kinematics implementations for controlling a $6$-DoF device using the $32$-bit fixed-point representation. The modules were implemented using $21$-bit for the fractional part and $11$-bit in the integer part. For the forward kinematics (FK), $3000 \, \text{ns}$ are required to perform all calculations, and for inverse kinematics (IK), $4500 \, \text{ns}$ is required. Based on the results of the implementations presented in this section, the implementation proposed for this work using floating-point representation had a speedup of $63.82\,\times$ for forward kinematics and $20.64\,\times$ for the inverse kinematics.

The research presented in \cite{2013_fpgaKinematicsCordic} proposed a hardware implementation of inverse kinematics to control a $10$-DoF device. The hardware was projected using the 32-bit fixed-point representation, however the amount of bits used in the fractional part was not specified. The architecture proposed to calculate the inverse kinematics requires $440 \, \text{ns}$ to perform the computation. Comparing to the inverse kinematics (IK) implementation using $32$-bit floating-point proposed by this work, the speedup was $2.01\,\times$ over the model presented in \cite{2013_fpgaKinematicsCordic}.

The authors in \cite{2015_fpgaKinematicArticuledRobot} present the results of fixed-point implementation for forward and inverse kinematics to control a $5$-DoF device, as described in Table \ref{tab:ResultsComparisons_FloatingPoint}. The proposed hardware implementation uses the numerical representation of $32$-bit ($15$-bit to fractional part) and $16$-bit ($7$-bit to fractional part) in different parts of the modules. The time required to perform the calculations is $680 \, \text{ns}$ and $940 \, \text{ns}$ for forward and inverse kinematics, respectively. Comparing to the floating-point implementation proposed by this work, the speedup was $14.46\,\times$ for forward kinematic and $4.31\,\times$ for inverse kinematic over the model presented in \cite{2015_fpgaKinematicArticuledRobot}.

Differently from previous works (Table \ref{tab:ResultsComparisons_FloatingPoint}), in \cite{2017_fpga_SurgicalRobotKinematic}, the authors present unique hardware for calculating forward and inverse kinematics together. In the proposed model, the 32-bit fixed-point representation was used. The total time to perform the calculation is $2000\, \text{ns}$. The time obtained was calculated taking into account the entire process duration, however,  separate times for each module were not specified. Given this scenario, by adding the $t_s$ FK module time that calculates forward kinematics with the IK module, the total time resulting from both implementations reaches $265\, \text{ns}$, according to Table \ref{tab:ResultsComparisons_FloatingPoint}. Hence, the hardware presented in the work here developed achieved a $7.54\,\times$ speedup over the model presented in \cite{2017_fpga_SurgicalRobotKinematic}.

It can be seen from Table \ref{tab:ResultsComparisons_FloatingPoint}, that none of the works from the state-of-the-art presented the hardware implementation of all four robotics algorithms that were presented here. It is also noted that just two works used the floating-point numerical representation. The floating-point implementation of robotics algorithms proposed by this work showed significant gains when compared to the works presented in the literature as shown in Table \ref{tab:ResultsComparisons_FloatingPoint}. The different amounts of degrees of freedom (DoF) used in the devices can somehow influence in values of sample rate and throughput. Another factor that can also influence these values is in relation to the type of FPGA that is used to perform the synthesis. Due to the fact that the implementation of this work was designed in a parallel architecture, the increase in the amount of DoF does not necessarily reflect in a significant increase in sample rate.

\section{Conclusions}

This paper presented a reconfigurable hardware reference model for four modules that implement robotics-associated algorithms. The FK-HMD and FK-HSD modules implement the forward kinematics, the IK-HSD module implements the inverse kinematics, the KFF-HMD module implements the kinesthetic feedback force and the FBF-HSM module implements the feedback force. The parallel FPGA implementation of the four modules is intended to increase the tactile system's processing speed with the purpose of meeting the latency constraints required for tactile internet applications. The modules were designed using a full-parallel implementation which works on a hybrid scheme that uses fixed point and floating point representation in distinct parts of the architecture.
Compared to the state of the art, this work stands out by presenting the description and implementation of four different robotics algorithms in FPGA. The implementations presented in this work achieve higher module processing speed when compared to equivalent implementations from the state-of-the-art. All of the modules here presented were analyzed based on the synthesis results, which included the hardware occupation area, sampling rate and throughput. Based on the synthesis results, it was observed that the implementations achieved high module processing speed, far below the latency limit of $1 \, \text{ms}$. Hardware modules achieved an acceleration of $93\,\times$ compared to the $37.5 \, \mu\text{s}$ time constraint. This demonstrates that using reconfigurable embedded systems on devices such as FPGAs enables parallel implementation of algorithms thus speeding up processing of the data and minimizing execution time. Runtime gains can make processing time possible for critical applications that require short time constraints or a large amount of data to be processed in a short time frame.

\section*{Acknowledgment}

This work was conducted during a scholarship supported by the Doctoral Sandwich Program CAPES/PDSE at the Federal University of Rio Grande do Norte. Financed by CAPES – Brazilian Federal Agency for Support and Evaluation of Graduate Education within the Ministry of Education of Brazil.

\bibliographystyle{IEEEtran}
\renewcommand{\refname}{References}
\bibliography{PaperMain}

\begin{thebibliography}{10}
\providecommand{\url}[1]{#1}
\csname url@samestyle\endcsname
\providecommand{\newblock}{\relax}
\providecommand{\bibinfo}[2]{#2}
\providecommand{\BIBentrySTDinterwordspacing}{\spaceskip=0pt\relax}
\providecommand{\BIBentryALTinterwordstretchfactor}{4}
\providecommand{\BIBentryALTinterwordspacing}{\spaceskip=\fontdimen2\font plus
\BIBentryALTinterwordstretchfactor\fontdimen3\font minus
  \fontdimen4\font\relax}
\providecommand{\BIBforeignlanguage}[2]{{%
\expandafter\ifx\csname l@#1\endcsname\relax
\typeout{** WARNING: IEEEtran.bst: No hyphenation pattern has been}%
\typeout{** loaded for the language `#1'. Using the pattern for}%
\typeout{** the default language instead.}%
\else
\language=\csname l@#1\endcsname
\fi
#2}}
\providecommand{\BIBdecl}{\relax}
\BIBdecl

\bibitem{tactile_basic2015_Mischa}
M.~Dohler, ``The tactile internet iot, 5g and cloud on steroids,'' in \emph{5G
  Radio Technology Seminar. Exploring Technical Challenges in the Emerging 5G
  Ecosystem}, March 2015, pp. 1--16.

\bibitem{tactile_network2015_RealizingTactile}
\BIBentryALTinterwordspacing
A.~Aijaz, M.~Dohler, A.~H. Aghvami, V.~Friderikos, and M.~Frodigh, ``Realizing
  the tactile internet: Haptic communications over next generation 5g cellular
  networks,'' \emph{CoRR}, vol. abs/1510.02826, 2015. [Online]. Available:
  \url{http://arxiv.org/abs/1510.02826}
\BIBentrySTDinterwordspacing

\bibitem{2017_challengesHapticTactile}
D.~V.~D. Berg, R.~Glans, D.~D. Koning, F.~A. Kuipers, J.~Lugtenburg,
  K.~Polachan, P.~T. Venkata, C.~Singh, B.~Turkovic, and B.~V. Wijk,
  ``Challenges in haptic communications over the tactile internet,'' \emph{IEEE
  Access}, vol.~5, pp. 23\,502--23\,518, 2017.

\bibitem{tactile_basic2016_Martin}
M.~Maier, M.~Chowdhury, B.~P. Rimal, and D.~P. Van, ``The tactile internet:
  vision, recent progress, and open challenges,'' \emph{IEEE Communications
  Magazine}, vol.~54, no.~5, pp. 138--145, 2016.

\bibitem{tactile_basic2016_The5G}
M.~Simsek, A.~Aijaz, M.~Dohler, J.~Sachs, and G.~Fettweis, ``The 5g-enabled
  tactile internet: Applications, requirements, and architecture,'' in
  \emph{2016 IEEE Wireless Communications and Networking Conference}, April
  2016, pp. 1--6.

\bibitem{REF_TIME_MS_01}
C.~Li, C.-P. Li, K.~Hosseini, S.~B. Lee, J.~Jiang, W.~Chen, G.~Horn, T.~Ji,
  J.~E. Smee, and J.~Li, ``5g-based systems design for tactile internet,''
  \emph{Proceedings of the IEEE}, vol. 107, no.~2, pp. 307--324, 2018.

\bibitem{8399482}
K.~{Antonakoglou}, X.~{Xu}, E.~{Steinbach}, T.~{Mahmoodi}, and M.~{Dohler},
  ``Toward haptic communications over the 5g tactile internet,'' \emph{IEEE
  Communications Surveys Tutorials}, vol.~20, no.~4, pp. 3034--3059,
  Fourthquarter 2018.

\bibitem{REF_TIME_MS_02}
A.~Nasrallah, A.~S. Thyagaturu, Z.~Alharbi, C.~Wang, X.~Shao, M.~Reisslein, and
  H.~ElBakoury, ``Ultra-low latency (ull) networks: The ieee tsn and ietf
  detnet standards and related 5g ull research,'' \emph{IEEE Communications
  Surveys \& Tutorials}, vol.~21, no.~1, pp. 88--145, 2018.

\bibitem{tactile_network2016_5GEnable}
M.~Simsek, A.~Aijaz, M.~Dohler, J.~Sachs, and G.~Fettweis, ``5g-enabled tactile
  internet,'' \emph{IEEE Journal on Selected Areas in Communications}, vol.~34,
  no.~3, pp. 460--473, 2016.

\bibitem{tactile_network2015_TowardsLatency}
D.~Szabo, A.~Gulyas, F.~H. Fitzek, F.~H. Fitzek, and D.~E. Lucani, ``Towards
  the tactile internet: Decreasing communication latency with network coding
  and software defined networking,'' in \emph{European Wireless 2015; 21th
  European Wireless Conference; Proceedings of}, May 2015, pp. 1--6.

\bibitem{internetSkills_mischa2017}
M.~Dohler, T.~Mahmoodi, M.~A. Lema, M.~Condoluci, F.~Sardis, K.~Antonakoglou,
  and H.~Aghvami, ``Internet of skills, where robotics meets ai, 5g and the
  tactile internet,'' in \emph{2017 European Conference on Networks and
  Communications (EuCNC)}, June 2017, pp. 1--5.

\bibitem{predicaoIAFPGA2015}
Q.~Yu, C.~Wang, X.~Ma, X.~Li, and X.~Zhou, ``A deep learning prediction process
  accelerator based fpga,'' in \emph{2015 15th IEEE/ACM International Symposium
  on Cluster, Cloud and Grid Computing}, May 2015, pp. 1159--1162.

\bibitem{marceloAlisson2014}
A.~C. de~Souza and M.~A. Fernandes, ``Parallel fixed point implementation of a
  radial basis function network in an fpga,'' \emph{Sensors}, vol.~14, no.~10,
  pp. 18\,223--18\,243, 2014.

\bibitem{8626462}
A.~L.~X. {Da Costa}, C.~A.~D. {Silva}, M.~F. {Torquato}, and M.~A.~C.
  {Fernandes}, ``Parallel implementation of particle swarm optimization on
  fpga,'' \emph{IEEE Transactions on Circuits and Systems II: Express Briefs},
  pp. 1--1, 2019.

\bibitem{8678408}
M.~G.~F. {Coutinho}, M.~F. {Torquato}, and M.~A.~C. {Fernandes}, ``Deep neural
  network hardware implementation based on stacked sparse autoencoder,''
  \emph{IEEE Access}, vol.~7, pp. 40\,674--40\,694, 2019.

\bibitem{Torquato2019}
M.~F. Torquato and M.~A.~C. Fernandes, ``High-performance parallel
  implementation of genetic algorithm on fpga,'' \emph{Circuits, Systems, and
  Signal Processing}, Jan 2019.

\bibitem{8574886}
L.~M.~D. {Da Silva}, M.~F. {Torquato}, and M.~A.~C. {Fernandes}, ``Parallel
  implementation of reinforcement learning q-learning technique for fpga,''
  \emph{IEEE Access}, vol.~7, pp. 2782--2798, 2019.

\bibitem{electronics8060631}
F.~F. Lopes, J.~C. Ferreira, and M.~A.~C. Fernandes, ``Parallel implementation
  on fpga of support vector machines using stochastic gradient descent,''
  \emph{Electronics}, vol.~8, no.~6, 2019.

\bibitem{NORONHA2019138}
D.~H. Noronha, M.~F. Torquato, and M.~A. Fernandes, ``A parallel implementation
  of sequential minimal optimization on fpga,'' \emph{Microprocessors and
  Microsystems}, vol.~69, pp. 138 -- 151, 2019.

\bibitem{2018_tactilePhysicalSystemDesign}
A.~N, A.~S. M, K.~Polachan, P.~T. V, and C.~Singh, ``An end to end tactile
  cyber physical system design,'' in \emph{2018 4th International Workshop on
  Emerging Ideas and Trends in the Engineering of Cyber-Physical Systems
  (EITEC)}, April 2018, pp. 9--16.

\bibitem{reduction_fpga2009_ImprovedHaptic}
M.~K. O’Malley, K.~S. Sevcik, and E.~Kopp, ``Improved haptic fidelity via
  reduced sampling period with an fpga-based real-time hardware platform,''
  \emph{Journal of Computing and Information Science in Engineering}, vol.~9,
  no.~1, p. 011002, 2009.

\bibitem{fpga_haptic2009_HapticComunnication}
H.~Tanaka, K.~Ohnishi, and H.~Nishi, ``Haptic communication system using fpga
  and real-time network framework,'' in \emph{Industrial Electronics, 2009.
  IECON'09. 35th Annual Conference of IEEE}.\hskip 1em plus 0.5em minus
  0.4em\relax IEEE, 2009, pp. 2931--2936.

\bibitem{reduction_fpga2013_AStudyFpga}
M.~Franc and A.~Hace, ``A study on the fpga implementation of the bilateral
  control algorithm towards haptic teleoperation,'' \emph{Automatika--Journal
  for Control, Measurement, Electronics, Computing and Communications},
  vol.~54, no.~1, 2013.

\bibitem{2010_fpgaKinematicRoboticManupulator}
D.~F. S{\'a}nchez, D.~M. Mu{\~n}oz, C.~H. Llanos, and J.~M. Motta, ``A
  reconfigurable system approach to the direct kinematics of a 5 dof robotic
  manipulator,'' \emph{International Journal of Reconfigurable Computing}, vol.
  2010, 2010.

\bibitem{2013_fpgaParallelRobotKinematic}
K.~Gac, G.~Karpiel, and M.~Petko, ``Fpga based hardware accelerator for
  calculations of the parallel robot inverse kinematics,'' in \emph{Proceedings
  of 2012 IEEE 17th International Conference on Emerging Technologies Factory
  Automation (ETFA 2012)}, Sept 2012, pp. 1--4.

\bibitem{fpga_2014_fixedPointKinematics}
M.~Wu, Y.~Kung, Y.~Huang, and T.~Jung, ``Fixed-point computation of robot
  kinematics in fpga,'' in \emph{2014 International Conference on Advanced
  Robotics and Intelligent Systems (ARIS)}, June 2014, pp. 35--40.

\bibitem{2013_fpgaKinematicsCordic}
C.~C. Wong and C.~C. Liu, ``Fpga realisation of inverse kinematics for biped
  robot based on cordic,'' \emph{Electronics Letters}, vol.~49, no.~5, pp.
  332--334, February 2013.

\bibitem{2015_fpgaKinematicArticuledRobot}
H.~Linh, B.~Thi, and Y.-S. Kung, ``Digital hardware realization of forward and
  inverse kinematics for a five-axis articulated robot arm,''
  \emph{Mathematical Problems in Engineering}, vol. 2015, 2015.

\bibitem{2017_fpga_SurgicalRobotKinematic}
Z.~Jiang, Y.~Dai, J.~Zhang, and S.~He, ``Kinematics calculation of minimally
  invasive surgical robot based on fpga,'' in \emph{2017 IEEE International
  Conference on Robotics and Biomimetics (ROBIO)}, Dec 2017, pp. 1726--1730.

\bibitem{phantom_guide}
Geomagic, \emph{\BIBforeignlanguage{English}{Phantom Omni, Device Guide}}.

\bibitem{2006_OmniPhantom_Teleoperation}
G.~Song, S.~Guo, and Q.~Wang, ``A tele-operation system based on haptic
  feedback,'' in \emph{2006 IEEE International Conference on Information
  Acquisition}, Aug 2006, pp. 1127--1131.

\bibitem{2012_OmniPhantom_Teleoperation}
T.~Sansanayuth, I.~Nilkhamhang, and K.~Tungpimolrat, ``Teleoperation with
  inverse dynamics control for phantom omni haptic device,'' in \emph{2012
  Proceedings of SICE Annual Conference (SICE)}, Aug 2012, pp. 2121--2126.

\bibitem{phantom2009_Phantom}
A.~J. Silva, O.~A.~D. Ramirez, V.~P. Vega, and J.~P.~O. Oliver, ``Phantom omni
  haptic device: Kinematic and manipulability,'' in \emph{Electronics, Robotics
  and Automotive Mechanics Conference, 2009. CERMA'09.}\hskip 1em plus 0.5em
  minus 0.4em\relax IEEE, 2009, pp. 193--198.

\bibitem{phantom2001_Kinematics}
M.~C. Cavusoglu and D.~Feygin, ``Kinematics and dynamics of phantom (tm) model
  1.5 haptic interface,'' 2001.

\bibitem{phantom2006_AStudy}
J.~San~Martin and G.~Trivi{\~n}o, ``A study of the manipulability of the
  phantom omni haptic interface.'' in \emph{VRIPHYS}, 2006, pp. 127--128.

\bibitem{8343385}
A.~{Kumar}, P.~J. {Gaidhane}, and V.~{Kumar}, ``A nonlinear fractional order
  pid controller applied to redundant robot manipulator,'' in \emph{2017 6th
  International Conference on Computer Applications In Electrical
  Engineering-Recent Advances (CERA)}, Oct 2017, pp. 527--532.

\bibitem{Yang2016}
C.~Yang, H.~Ma, and M.~Fu, \emph{Intelligent Control of Robot
  Manipulator}.\hskip 1em plus 0.5em minus 0.4em\relax Singapore: Springer
  Singapore, 2016, pp. 49--96.

\bibitem{Nazemizadeh2014}
H.~Rahimi and M.~Nazemizadeh, ``Dynamic analysis and intelligent control
  techniques for flexible manipulators: a review,'' \emph{Advanced Robotics},
  vol.~28, no.~2, pp. 63--76, 2014.

\bibitem{SaiHong2014}
S.~H. Tang, C.~K. Ang, M.~K. A. B.~M. Ariffin, and S.~B. Mashohor, ``Predicting
  the motion of a robot manipulator with unknown trajectories based on an
  artificial neural network,'' \emph{International Journal of Advanced Robotic
  Systems}, vol.~11, no.~10, p. 176, 2014.

\bibitem{app8122648}
\BIBentryALTinterwordspacing
Y.~Chen and L.~Li, ``Predictable trajectory planning of industrial robots with
  constraints,'' \emph{Applied Sciences}, vol.~8, no.~12, 2018. [Online].
  Available: \url{https://www.mdpi.com/2076-3417/8/12/2648}
\BIBentrySTDinterwordspacing

\bibitem{Xiang2019}
Y.~Xiang, ``Simulation and analysis of three-dimensional space path prediction
  for six-degree-of-freedom (sdof) manipulator,'' \emph{3D Research}, vol.~10,
  no.~2, p.~15, Apr 2019.

\bibitem{6094666}
B.~{Bócsi}, D.~{Nguyen-Tuong}, L.~{Csató}, B.~{Schölkopf}, and J.~{Peters},
  ``Learning inverse kinematics with structured prediction,'' in \emph{2011
  IEEE/RSJ International Conference on Intelligent Robots and Systems}, Sep.
  2011, pp. 698--703.

\bibitem{electronics8040398}
S.~Shen, A.~Song, and T.~Li, ``Predictor-based motion tracking control for
  cloud robotic systems with delayed measurements,'' \emph{Electronics},
  vol.~8, no.~4, 2019.

\bibitem{FeedbackForce2018}
C.~Yang, Y.~Xie, S.~Liu, and D.~Sun, ``Force modeling, identification, and
  feedback control of robot-assisted needle insertion: A survey of the
  literature,'' in \emph{Sensors}, 2018.

\bibitem{s19225029}
\BIBentryALTinterwordspacing
J.~C. V.~S. Junior, M.~F. Torquato, D.~H. Noronha, S.~N. Silva, and M.~A.~C.
  Fernandes, ``Proposal of the tactile glove device,'' \emph{Sensors}, vol.~19,
  no.~22, 2019. [Online]. Available:
  \url{https://www.mdpi.com/1424-8220/19/22/5029}
\BIBentrySTDinterwordspacing

\bibitem{REF20_RelatedWork04}
P.~Weber, E.~Rueckert, R.~Calandra, J.~Peters, and P.~Beckerle, ``A low-cost
  sensor glove with vibrotactile feedback and multiple finger joint and hand
  motion sensing for human-robot interaction,'' in \emph{2016 25th IEEE
  International Symposium on Robot and Human Interactive Communication
  (RO-MAN)}.\hskip 1em plus 0.5em minus 0.4em\relax IEEE, 2016, pp. 99--104.

\bibitem{REF19_RelatedWork03}
N.~Arjun, S.~Ashwin, K.~Polachan, T.~Prabhakar, and C.~Singh, ``An end to end
  tactile cyber physical system design,'' in \emph{2018 4th International
  Workshop on Emerging Ideas and Trends in the Engineering of Cyber-Physical
  Systems (EITEC)}.\hskip 1em plus 0.5em minus 0.4em\relax IEEE, 2018, pp.
  9--16.

\bibitem{algoritmoCordic}
J.~E. Volder, ``The cordic trigonometric computing technique,'' \emph{IRE
  Transactions on Electronic Computers}, no.~3, pp. 330--334, 1959.

\end{thebibliography}

\EOD

\end{document}